\documentclass[a4paper,prd,showpacs,preprintnumbers,superscriptaddress,nofootinbib,preprint]{revtex4-1}

\usepackage[dvipdfmx]{graphicx}
\usepackage{dcolumn}
\usepackage{bm}
\usepackage{hyperref}
\usepackage{color}
\usepackage{cancel}
\usepackage{amsmath,amsfonts,amssymb}
\usepackage{setspace}

\begin{document}


\def\a{\alpha}
\def\b{\beta}
\def\c{\varepsilon}
\def\d{{\mit\D}}
\def\e{\epsilon}
\def\f{\phi}
\def\g{\gamma}
\def\h{\theta}
\def\k{\kappa}
\def\l{\lambda}
\def\m{\mu}
\def\n{\nu}
\def\p{\psi}
\def\q{\partial}
\def\r{\rho}
\def\s{\sigma}
\def\t{\tau}
\def\u{\upsilon}
\def\v{\varphi}
\def\w{\omega}
\def\x{\xi}
\def\y{\eta}
\def\z{\zeta}
\def\D{\Delta}
\def\G{\Gamma}
\def\H{\Theta}
\def\L{\Lambda}
\def\F{\Phi}
\def\P{\Psi}
\def\S{\Sigma}

\def\o{\over}
\newcommand{\gsim}{ \mathop{}_{\textstyle \sim}^{\textstyle >} }
\newcommand{\lsim}{ \mathop{}_{\textstyle \sim}^{\textstyle <} }
\newcommand{\vev}[1]{ \left\langle {#1} \right\rangle }
\newcommand{\bra}[1]{ \langle {#1} | }
\newcommand{\ket}[1]{ | {#1} \rangle }
\newcommand{\EV}{ {\rm eV} }
\newcommand{\KEV}{ {\rm keV} }
\newcommand{\MEV}{ {\rm MeV} }
\newcommand{\GEV}{ {\rm GeV} }
\newcommand{\TEV}{ {\rm TeV} }
\newcommand{\1}{\mbox{1}\hspace{-0.25em}\mbox{l}}
\newcommand{\headline}[1]{\noindent{\bf #1}}
\def\diag{\mathop{\rm diag}\nolimits}
\def\Spin{\mathop{\rm Spin}}
\def\SO{\mathop{\rm SO}}
\def\O{\mathop{\rm O}}
\def\SU{\mathop{\rm SU}}
\def\U{\mathop{\rm U}}
\def\Sp{\mathop{\rm Sp}}
\def\SL{\mathop{\rm SL}}
\def\tr{\mathop{\rm tr}}
\def\mpl{M_{PL}}

\def\IJMP{Int.~J.~Mod.~Phys. }
\def\MPL{Mod.~Phys.~Lett. }
\def\NP{Nucl.~Phys. }
\def\PL{Phys.~Lett. }
\def\PR{Phys.~Rev. }
\def\PRL{Phys.~Rev.~Lett. }
\def\PTP{Prog.~Theor.~Phys. }
\def\ZP{Z.~Phys. }

\def\dd{\mathrm{d}}
\def\ff{\mathrm{f}}
\def\BH{{\rm BH}}
\def\inf{{\rm inf}}
\def\ev{{\rm evap}}
\def\eq{{\rm eq}}
\def\SM{{\rm sm}}
\def\Mpl{M_{\rm Pl}}
\def\GeV{{\rm GeV}}
\newcommand{\Red}[1]{\textcolor{red}{#1}}


\preprint{IPMU19-0002}

\title{Decay of I-ball/Oscillon in Classical Field Theory}
\author{Masahiro Ibe}
\email[e-mail: ]{ibe@icrr.u-tokyo.ac.jp}
\affiliation{ICRR, The University of Tokyo, Kashiwa, Chiba 277-8582, Japan}
\affiliation{Kavli IPMU (WPI), UTIAS, The University of Tokyo, Kashiwa, Chiba 277-8583, Japan}
\author{Masahiro Kawasaki}
\email[e-mail: ]{kawasaki@icrr.u-tokyo.ac.jp}
\affiliation{ICRR, The University of Tokyo, Kashiwa, Chiba 277-8582, Japan}
\affiliation{Kavli IPMU (WPI), UTIAS, The University of Tokyo, Kashiwa, Chiba 277-8583, Japan}
\author{Wakutaka Nakano}
\email[e-mail: ]{m156077@icrr.u-tokyo.ac.jp}
\affiliation{ICRR, The University of Tokyo, Kashiwa, Chiba 277-8582, Japan}
\affiliation{Kavli IPMU (WPI), UTIAS, The University of Tokyo, Kashiwa, Chiba 277-8583, Japan}
\author{Eisuke Sonomoto}
\email[e-mail: ]{sonomoto@icrr.u-tokyo.ac.jp}
\affiliation{ICRR, The University of Tokyo, Kashiwa, Chiba 277-8582, Japan}
\affiliation{Kavli IPMU (WPI), UTIAS, The University of Tokyo, Kashiwa, Chiba 277-8583, Japan}


\begin{abstract}
I-balls/oscillons are long-lived and spatially localized solutions of real scalar fields.
They are produced in various contexts of the early universe in, such as, the inflaton evolution and the axion evolution.
However, their decay process has long been unclear.
In this paper, we derive an analytic formula of the decay rate of the I-balls/oscillons within the classical field theory.
In our approach, we calculate the Poynting vector of the perturbation around the I-ball/oscillon profile
by solving a relativistic field equation, with which the decay rate of the I-ball/oscillon is obtained.
We also perform a classical lattice simulation and confirm the validity of our analytical formula of the decay rate numerically.
\end{abstract}

\maketitle


\section{Introduction}
Scalar fields are essential ingredients in particle physics and cosmology.
They are ubiquitous in many low energy effective field theories,
as they provide concise descriptions of spontaneous symmetry breaking.
The scalar fields corresponding to the Nambu-Goldstone bosons also appear in many well-motivated high energy theories.
In the de facto standard model of the cosmic inflation, inflation is driven by the scalar potential of a scalar field, the inflaton~\cite{Guth:1980zm,Linde:1981mu,Albrecht:1982wi,Sato:1980yn}.
The scalar fields are also indispensable if supersymmetry is realized in nature.

In this paper, we study the time-evolution of the I-ball/oscillon which appears in real scalar field theories.
The I-ball/oscillon has long been recognized as a spatially localized solitonic state which appears in a real scalar field theory~\cite{Bogolyubsky:1976yu,Gleiser:1993pt,Copeland:1995fq}.
The I-ball/oscillon associates with the conserved charge, the adiabatic charge $I$~\cite{Kasuya:2002zs,Kawasaki:2015vga},
as the topological solitons (i.e., domain walls, monopoles, cosmic strings)~\cite{Zeldovich:1974uw,tHooft:1974kcl,Polyakov:1974ek,Kibble:1976sj}
as well as the non-topological solitons (i.e., Q-balls)~\cite{Coleman:1985ki,Kusenko:1997si,Enqvist:1997si,Enqvist:1998en,Kasuya:1999wu,Kasuya:2000wx}
associate with their corresponding conserved (topological) charges.
The I-ball/oscillon can also be regarded as a Q-ball in the non-relativistic field theory where the adiabatic charge $I$
is reduced to the charge of an approximate $U(1)$ symmetry related to the particle number conservation~\cite{Mukaida:2014oza}.
These two pictures are consistent with each other since the adiabatic invariant $I$ is well conserved when the quadratic potential dominates the scalar potential, and hence when the non-relativistic limit is valid.

The  I-ball/oscillon is produced in various contexts of the early universe.
For example, the oscillations of the inflaton after inflation can lead to a strong inhomogeneity through the self-resonance,
which results in the formation of the I-ball/oscillon~\cite{McDonald:2001iv,Amin:2010jq,Amin:2011hj,Amin:2013ika,Takeda:2014qma,Lozanov:2016hid,Hasegawa:2017iay,Antusch:2017flz,Hong:2017ooe}.
The inflatonic I-ball/oscillon formation produces the gravitational wave, and its spectrum is studied in Ref.~\cite{Zhou:2013tsa,Antusch:2016con}.
The axion can also form the I-ball/oscillon which is sometime
called ``axiton"\cite{Kolb:1993zz,Kolb:1993hw,Visinelli:2017ooc,Vaquero:2018tib}.
The axion~\cite{Weinberg:1977ma,Wilczek:1977pj,Kim:1979if,Shifman:1979if,Dine:1981rt}  is the Nambu-Goldstone boson associated with spontaneous symmetry breaking of the Peccei-Quinn symmetry~\cite{Peccei:1977ur,Peccei:1977hh},
which is the most attractive solution to the strong CP problem~\cite{tHooft:1976rip}.
Due to the axiton formation, the axion can be spatially localized in the universe, which could have a significant impact on the axion search experiments.

The conservation of the adiabatic charge $I$ or the $U(1)$ charge in the non-relativistic limit
is not exact.
Accordingly, the I-ball/oscillon is not completely stable and decays eventually.
Although physics of the I-ball/oscillon have been studied in many
papers~\cite{Kawasaki:2015vga,Fodor:2006zs,Fodor:2008du,Gleiser:2008ty,Fodor:2009kf,Gleiser:2009ys,Hertzberg:2010yz,Saffin:2014yka,Kawasaki:2013awa},
the decay process  of the I-ball/oscillon has not been fully understood.
It is only recently that an analytic formula of the I-ball/oscillon decay has been derived based on the Q-ball picture where the decay rate is calculated 
in the Feynman diagrammatic approach~\cite{Mukaida:2016hwd,Eby:2018ufi}.

The main purpose of this paper is to revisit the decay process of the I-ball/oscillon. In our approach, we solve the relativistic
classical field equation of the perturbation around the I-ball/oscillon solution.
By calculating the Poynting vector of the perturbation,
we estimate how the localized energy of the I-ball/oscillon leaks out, which gives the decay rate of the I-ball/oscillon.
Because our analysis only uses the classical field equation, it is more straightforward than the analysis in~\cite{Mukaida:2016hwd,Eby:2018ufi}.
Our analysis also clarifies the physical picture of the I-ball/oscillon decay.
The decay process is just a leakage of the localized energy of the I-ball/oscillon via a classical emission of the relativistic modes of the scalar field.
We also validate our analytical formula of the decay rate by performing a classical lattice simulation.

Organiztion of this paper is as follows.
In Sec.~\ref{sec:th}, we briefly review the I-ball/oscillon solution in both the Q-ball picture and the adiabatic invariant picture.
In Sec.~\ref{sec:decay}, we calculate the I-ball/oscillon decay rate by solving a relativistic field equation of the perturbation around the I-ball/oscillon configuration.
In Sec.~\ref{sec:num_sim}, we perform a classical lattice simulation to validate our perturbative analysis.
Finally in Sec. \ref{sec:con_dis}, we summarize our results.

\section{I-ball/Oscillon Solution}
\label{sec:th}
In this section, we briefly review the I-ball/oscillon solution in a real scalar field theory.
In \ref{sec:I-Q} we explain the Q-ball description of the I-ball/oscillon following Ref.~\cite{Mukaida:2014oza,Mukaida:2016hwd}
and see that the I-ball/oscillon associates with the particle number conservation.
In \ref{sec:I-ad}, we re-derive the I-ball/oscillon solution by using the conservation of the adiabatic charge~\cite{Kasuya:2002zs,Kawasaki:2015vga}.
The I-ball/oscillon profiles derived by these two approaches coincide with each other
when the quadratic potential dominates its scalar potential.

\subsection{I-ball/oscillon  as Q-ball}
\label{sec:I-Q}

Let us consider a classical field theory of a real scalar field $\phi$ with a Lagrangian density

\newpage

\begin{eqnarray}
    {\cal L} = \frac{1}{2}\q^\mu\phi(x) \q_\mu \phi(x) - \frac{1}{2}m^2 \phi^2  - V(\phi)\ ,
\end{eqnarray}
where we assume a scalar potential with coupling constants $g_n$ as
\begin{eqnarray}
\label{eq:V1}
    V(\phi) = \sum_{n\ge 3} \frac{g_n}{n!} \phi^n\ .
\end{eqnarray}
The equation of motion of the field is represented by
\begin{eqnarray}
    \label{eq:EOM1}
    \left(\Box + m^2\right) \phi(x) = - V',
\end{eqnarray}
where $V' = \q V/\q \phi$ and the corresponding energy density is
\begin{eqnarray}
     {\cal E} = \frac{1}{2}\dot\phi(x)^2
     + \frac{1}{2} \left( \nabla\phi(x) \right)^2
     + \frac{1}{2}m^2 \phi^2  + V(\phi)\ .
\end{eqnarray}

Let us take the non-relativistic limit by expanding $\phi(x)$ by a complex scalar field $\Psi$;
\begin{eqnarray}
    \phi_\text{NR}(t,{\mathbf x})
       = \Psi(t,\mathbf x)e^{-imt} + \Psi(t,\mathbf x)^\dagger e^{imt}\ ,
\end{eqnarray}
where we assume $|\q_0 \Psi| \ll |m\Psi|$,  $|\q_0^2 \Psi| \ll |m^2\Psi|$,
and  $|\nabla^2 \Psi| \ll |m^2\Psi|$.
By substituting $\phi_\text{NR}$ to the Lagrangian and the energy density and taking time average of them with a time scale much longer than $m^{-1}$ but much shorter than that of the time variation of $\Psi(t,{\mathbf x})$,
the terms proportional to $e^{i n m t} (n\neq 0)$ drop out.
The resultant effective Lagrangian and the time-averaged energy density are represented by%
\footnote{Here, the time-averaged energy density $\overline{\cal E}$ does not coincide with
the effective Hamiltonian density derived from the effective Lagrangian in Eq.\,(\ref{eq:effL}),
\begin{eqnarray}
\label{eq:NRH}
{\cal H}_{\rm NR} = |\dot \Psi|^2 + |\nabla \Psi|^2 + V_{\rm eff}\ ,
\end{eqnarray}
with $\Pi = \dot\Psi - i m \Psi$ being the canonical momentum of $\Psi^\dagger$.
}
\begin{eqnarray}
\label{eq:effL}
    {\cal L}_{\rm NR}
        &=& \q^{\mu} \Psi \q_{\mu} \Psi^\dagger - im\Psi(\q_0 \Psi^\dagger)
            + im(\q_0\Psi) \Psi^{\dagger}
            - V_{\rm eff}  \ ,\\[0.5em]
            \label{eq:effE}
    \overline{\cal E}
        &=& |\dot \Psi -i m \Psi|^2+ |\nabla \Psi|^2 + m^2 |\Psi|^2
        + V_{\rm eff}\ , \\[0.5em]
        \label{eq:Veff1}
    V_{\rm eff}
        &=& \sum_{n \ge 2} \frac{g_{2n}}{(2n)!}{}_{2n}C_{n} (\Psi\Psi^{\dagger})^{n}
         =  \sum_{n \ge 2} \frac{g_{2n}}{(n!)^2} (\Psi\Psi^{\dagger})^{n}     \ .
\end{eqnarray}

In this approximation terms with the odd number of $\Psi$ vanish and the time averaged Lagrangian shows a $U(1)$ symmetry which corresponds to the conservation of the particle number.
The conserved charge is represented by,\footnote{
    The corresponding symmetry is $\Psi \to \Psi' = e^{i \alpha} \Psi$.
}
\begin{eqnarray}
    Q = - i\int  d^3x
        \left(\Psi (\dot\Psi-im\Psi)^\dagger
           -  \left(\dot\Psi-im\Psi\right)\Psi^\dagger\right)\ .
\end{eqnarray}
It should be stressed that no particle creation is allowed via the interaction terms in the non-relativistic limit,
which is the reason why we have an approximate $U(1)$ symmetry.

Now, let us find a Q-ball solution for a given $Q_0$ by the Lagrangian multiplier method because the field configuration of the Q-ball is obtained by minimizing the time-averaged energy for 
a given charge.
\begin{eqnarray}
    I_{Q_0} &=& \int d^3x\, \overline{\cal E} + \omega (Q_0-Q) \\
            &=& \int d^3x \left[
                |\dot \Psi -i (m-\omega) \Psi|^2+ \nabla \Psi \nabla \Psi^\dagger
                + (m^2-\omega^2) |\Psi|^2+   V_{\rm eff}\right]
            + \omega Q_0\ .
\end{eqnarray}
Then, a Q-ball solution
\begin{eqnarray}
    \Psi_Q &=& e^{i\mu t} \psi(r)\ , \quad(\psi \in {\mathbb R})\ , \\[0.5em]
    \mu &=& m-\omega\ , \quad(\mu \ll m \in {\mathbb R})\ ,
\end{eqnarray}
should satisfy
\begin{eqnarray}
    \label{eq:EOM2}
    \left[\frac{\q^2}{\q r^2}+ \frac{2}{r} \frac{\q}{\q r}\right] \psi(r)
    &=&  (2 \mu m - \mu^2) \psi(r) +
     \frac{1}{2} V_{\rm eff}' (\psi) \ ,
    \\
    \lim_{r\to 0} \frac{\q \psi(r)}{\q r}
    &=& \lim_{r \to \infty} \psi(r) = 0\ .
\end{eqnarray}
where
\begin{eqnarray}
    V_{\rm eff}(\psi) = \sum_{n \ge 2} \frac{g_{2n}}{(n!)^2} \psi^{2n} \ .
\end{eqnarray}
and $V_{\rm eff}'(\psi)$ denotes the derivative with respect to $\psi$.%
\footnote{It should be noted that $\partial V_{\rm eff}(\Psi\Psi^\dagger)/\partial\Psi = \partial V_{\rm eff}(\Psi\Psi^\dagger)/\partial \Psi^\dagger = \partial V_{\rm eff}(\psi)/\partial \psi/2$.
}

The necessary condition for the existence of solutions of Eq.~(\ref{eq:EOM2}) is
\begin{eqnarray}
    0  < 2\m m - \mu^2 < - \min\left[ V_{\rm eff}/\psi^2 \right]\ ,
    \label{oscillon_condition}
\end{eqnarray}

The parameter $\omega = m-\mu$ is chosen so that the solution satisfies
\begin{eqnarray}
    Q_0=8\pi \omega \int  dr\, r^2 \psi(r)^2\ .
\end{eqnarray}
The total energy of the solution is given by
\begin{eqnarray}
    E =  4\pi \int dr \,r^2 \left[
      (\omega^2 +m^2)\psi^2(r)
      + \left(\q_r \psi(r)\right)^2
      +{V}_{\rm eff}(\psi(r))
      \right]\ .
\end{eqnarray}
With these definitions, we can show
\begin{eqnarray}
    \frac{dE}{dQ_0} = \omega \ ,
    \label{eq:e-q}
\end{eqnarray}
by taking derivative of $\omega$ and using the equation of motion Eq.\,(\ref{eq:EOM2}).%

Finally, let us comment on the relation between the time-averaged energy density Eq.\,(\ref{eq:effE})
and the Hamiltonian density Eq.\,(\ref{eq:NRH}).
For the I-ball/oscillon solution, these densities are related via
\begin{eqnarray}
\overline{\cal E} = {\cal H}_{\rm NR} + \frac{m^2-2\mu m}{2(m-\mu)} q_0\ ,
\end{eqnarray}
where $q_0$ is the charge density of the I-ball/oscillon i.e. $q_0 = 2(m-\mu) \psi^2(r)$.
Thus, the I-ball/oscillon solution which minimizes $\overline{\cal E}$ for a given value $Q_0$
also minimizes ${\cal H}_{\rm NR}$.

\subsection{I-ball/oscillon from Adiabatic Invariance}
\label{sec:I-ad}

The I-ball/oscillon solutions are obtained in Ref.~\cite{Kasuya:2002zs} as localized scalar field configurations which minimize their time-averaged energy for a given adiabatic charge $I$.
The adiabatic invariant approximately conserves when the scalar field dynamics is dominated by a quadratic potential.%
\footnote{
It is shown in Ref.\,\cite{Kawasaki:2015vga} that, only for the particular potential, $I$ is exactly conserved and the oscillon is expected to be stable classically.}

The adiabatic invariance is defined as
\begin{equation}
\label{eq:I}
    I = \frac{1}{\omega} \int d^3 x \overline{\dot{\phi}^2} \ ,
\end{equation}
where $\omega$ is the angular frequency of the oscillating field and the overbar denotes the average over one period of the oscillation.
The I-ball/oscillon solution is obtained by minimizing
\begin{equation}
    E_{\lambda} = \int d^3 x \left( \frac{1}{2} \overline{\dot{\phi}^2}
        + \frac{1}{2} \overline{(\nabla \phi)^2} + \frac{1}{2} m^2 \overline{\phi^2}
        + \overline{V} \right)
        + \lambda \left(I_0 -I\right)
\end{equation}
where $\lambda$ is the Lagrange multiplier and $V$ denotes the scalar potential in Eq.\,(\ref{eq:V1}).
Since the I-ball/oscillon solution exists when the mass term dominates the scalar potential, the solution can be written as $\phi (t, {\bf x}) = 2 \psi ({\bf x}) \cos (\omega t)$ in good approximation, where $\omega$ is nearly equal to but less than $m$.
Thus we define $\mu$ as $\mu = m - \omega \ll m$.

Using $\phi = 2 \psi \cos (\omega t)$, $E_{\lambda}$ is rewritten as
\begin{equation}
    E_{\lambda} = \int d^3 x
        \left[  (\nabla \psi)^2 + ( m^2 + \omega^2
          - 2 \lambda \omega ) \psi^2 + V_{\mathrm{eff}} (\psi) \right]
        + \lambda I_0 \ ,
\end{equation}
where the-averaged scalar potential $\overline{V}$ coincides with $V_{\rm eff}(\psi)$ in Eq.\,(\ref{eq:Veff1}).
Assuming the configuration is spherical, i.e. $\psi ({\bf x}) = \psi (r)$, the I-ball/oscillon solution is obtained from
\begin{equation}
    \label{eq:I-ball eom}
    \left[
        \frac{\q^2}{\q r^2}+ \frac{2}{r} \frac{\q}{\q r}
    \right] \psi(r) =  (m^2 + \omega^2 - 2 \omega \lambda) \psi(r)    + \frac{1}{2} \frac{d V_{\rm eff}}{d \psi} (\psi)\ ,
\end{equation}
with the boundary condition,
\begin{equation}
    \lim_{r\to 0} \frac{\q \psi(r)}{\q r} = \lim_{r \to \infty} \psi(r) = 0\ .
\end{equation}

The Lagrange multiplier $\lambda$ is determined by using equation of motion for $\phi$ which is given by
\begin{equation}
    \ddot{\phi} - \nabla^2 \phi + m^2 \phi + \frac{d V}{d \phi} (\phi) = 0 \ .
\end{equation}
Substituting $\phi = 2 \psi \cos (\omega t)$,
\begin{equation}
    - \omega^2 \psi \cos (\omega t)
    - \nabla^2 \psi \cos (\omega t)
    + m^2 \psi \cos (\omega t)
    + \frac{1}{2} \frac{d V}{d \phi} (2 \psi \cos (\omega t)) = 0 \ .
\end{equation}
Multiplying this equation by $\cos (\omega t)$ and averaging over a period, we obtain
\begin{equation}
    \label{eq:I-ball eom2}
    \nabla^2 \psi = (m^2 - \omega^2) \psi
         + \frac{1}{2} \frac{d V_{\rm eff}}{d \psi} (\psi) \ .
\end{equation}
Comparing Eq.\,(\ref{eq:I-ball eom}) and Eq.\,(\ref{eq:I-ball eom2}), we find

\newpage

\begin{equation}
    \lambda = \omega = m - \mu \ .
\end{equation}
It means that $\omega$ is chosen to minimize the adiabatic invariant $I_0$.

As a result, we see that the I-ball/oscillon solution associates with the adiabatic charge $I_0$ is the same with the one derived in the previous section.
The correspondence between the two approach is more evident by noting that the $U(1)$ charge $Q$ is nothing but
the adiabatic charge
\begin{eqnarray}
Q_0 = I_0 = 8\pi \omega \int  dr\, r^2 \psi(r)^2\ ,
\end{eqnarray}
for the I-ball/oscillon solution.
It should be again emphasized that the conservation of the adiabatic charge and the approximate $U(1)$ charge are
valid when the scalar potential is dominated by the quadratic term, which makes the oscillation frequency of the real scalar
field very close to $m$, i.e. $\mu \ll m$.

\section{Analytical calculation of I-ball/oscillon decay}
\label{sec:decay}
In this section, we derive a formula of the scalar radiation from I-ball/oscillon in the classical field theory.
For a given I-ball/oscillon solution, we solve the equation of motion of the perturbation
and calculate the energy loss rate of the I-ball/oscillon.

\subsection{Scalar Radiation from I-ball/oscillon}
\label{Scalar_radiation}
Let us take the I-ball/oscillon (Q-ball) solution at $t = t_0$ and consider the perturbation $\xi$ around it,
\begin{eqnarray}
    \phi(x) =2\psi(r)\cos(\omega t)+ \xi(x)\ ,
\end{eqnarray}
with $\omega  = m-\mu$.
The I-ball/oscillon solution constructed in the previous section satisfies%
\begin{eqnarray}
\label{eq:I-ballEQ}
    (\Box + m^2 ) 2\psi(r)\cos\omega t = - V_{\rm eff}'(\psi) \cos\omega t\ .
\end{eqnarray}

When the perturbation is small, i.e., $|\xi|\ll |\psi|$, the right-hand side of the equation of motion in Eq.\,(\ref{eq:EOM1}) can be approximated by
\vspace{-0.1cm}
\begin{eqnarray}
\label{eq:Vprime}
    V'(\phi)  \simeq V'(2\psi(r)\cos \omega t)
        =  \sum_{n\ge 3} \frac{2^{n-1}g_n}{(n-1)!} \psi^{n-1} \cos^{n-1}\omega t \ .
\end{eqnarray}
\vspace{-0.1cm}
In this approximation, the back reaction of the radiation is neglected.
The equation of motion of $\xi$ is written as
\begin{eqnarray}
    (\Box + m^2)\xi &=&  - \sum_{n\ge 1} \rho_n(r)\cos^n\omega t \ , \\
    \rho_1(r) &=&  -\sum_{\ell\ge 2} \frac{2\ell g_{2\ell}}{(\ell!)^2 }\psi(r)^{2\ell-1} \ , \label{eq:rho_1}\\
    \rho_n(r) &=&\frac{2^{n}g_{n+1}}{(n)!} \psi(r)^{n} \ , \quad (n\ge 2) \ .
    \label{eq:wave}
\end{eqnarray}
Here, $\rho_1$ denotes the contribution from the right-hand side of Eq.\,(\ref{eq:I-ballEQ}),
while $\rho_n$'s come from the right-hand size of Eq.\,(\ref{eq:Vprime}).
Using $\cos^n \omega t = \sum_{k = 0}^{n}\frac{1}{2^n} {}_nC_k \cos((n-2k) \omega t)$, we may further reduce the source term to
\begin{eqnarray}
    (\Box + m^2)\xi  &=&
    - \sum_{n\ge 1}\sum_{k = 0}^{n} \frac{1}{2^n} {}_nC_k \rho_n(r)\cos ((n-2k)\omega t)\  .
    \label{eq:wave2}
\end{eqnarray}
As we will see shortly, $\rho _1 (r)$ in Eq.~(\ref{eq:rho_1}) does not contribute to the scalar radiation.

To solve the equation of motion of $\xi$, let us assume that the I-ball/oscillon is placed at $t_0\to -\infty$, so that $\xi$  is radiated constantly.
In this setup, the equation of motion can be easily solved by using the Fourier transformed fields,
\begin{eqnarray}
    \hat\xi(p^0 ,{\mathbf p})
      &=& \int d^4x\,  \xi(t,{\mathbf x})
          e^{ip^0 t-i{\mathbf p}\cdot {\mathbf x}} \ ,\\
    \hat\rho(p^0 ,{\mathbf p})
      &=& \int d^4x\,  \rho(t,{\mathbf x})
          e^{ip^0 t-i{\mathbf p}\cdot {\mathbf x}}  \ ,\\
          \label{eq:retG}
    \hat G(p^0,{\mathbf p})
      &=& \int d^4 x\,  G_{\rm ret}(t,{\mathbf x})
          e^{ip^0 t-i{\mathbf p}\cdot {\mathbf x}}  \\
      &=& \lim_{\varepsilon \to +0}
          \frac{1}{(p^0 +i\varepsilon )^2  - {\mathbf p}^2 - m^2}\ ,
\end{eqnarray}
where $G_\text{ret}$ is the Green function satsifying $(\Box + m^2)G_\text{ret}=\delta(x)$ 
with the retarded boundary condition, i.e. $\varepsilon > 0$.
Here, $\rho(t,{\mathbf x})$ denotes the right-hand side  ($\times (-1)$) of Eq.\,(\ref{eq:wave2}).
It should be noticed that the source at $t'$ only affects $\xi(t)$ for $t > t'$.
The domain of $p^0$ is $(-\infty, \infty)$ as it just parameterizes the frequency.
By using the Fourier transformation of $\rho(t,{\mathbf x})$ in Eq.\,(\ref{eq:wave2}),
$\hat{\rho}(p^0,\mathbf{x})$ is written as
\begin{eqnarray}
    \hat\rho(p^0,{\mathbf p})
        &=& \sum_{n\ge 1} \sum_{k = 0}^{n}\frac{\pi}{2^n} {}_nC_k \left(
            \delta(p_0 -(n-2k) \omega )
            + \delta(p_0 +(n-2k) \omega )\right)\tilde\rho_n(p) \ ,\\
    \tilde\rho_n(p)
        & = & \int d^3 x\,  \rho_n(r) e^{i{\mathbf p}\cdot {\mathbf x} }
          = 4\pi\int dr  \rho_n(r)  \frac{r \sin pr}{p} \ , \\
         \tilde \rho_n(-p) &=& \tilde \rho_n (p) \ .
    \label{eq:source}
\end{eqnarray}
Thus, $\hat\rho(p^0,{\mathbf p}) $ does not depend on the direction of ${\mathbf p}$ but only on $p = |{\mathbf p}|$.
Solving the equation of motion of $\xi$ (see the appendix\,\ref{sec:calculation} for a detailed derivation), we obtain
\begin{eqnarray}
\label{eq:ASsol}
   \xi(t,{\mathbf x})
      &=&     \sum_{n\ge 2} \sum_{\{k| \omega_{nk}>m\}}^n
         \frac{-1}{(2\pi)}
         \frac{g_{n+1}}{k!(n-k)!} \tilde\psi_n(\overline{\omega}_{nk} ) \,
         \frac{1}{r} \cos(\omega_{nk}t -     \overline{\omega}_{nk}r)\ , \\
   \tilde\psi_{n}(p)
      &=& 4\pi\int dr\,  \psi^n(r)  \frac{r \sin pr}{p}\ ,\\
   \omega_{nk}
      &=& |(n-2k)\omega| = |(n-2k)(m - \mu) |  > m \ , \\
     \overline{\omega}_{nk} &=&(\omega_{nk}^2-m^2)^{1/2} \ ,
\end{eqnarray}
for $r\to \infty$.
Here, the summation over $k$ is taken only for $\omega_{nk} > m$, and hence, $n = 1$ does not contribute since $ m - \mu < m$.
Therefore, $\rho_1$, and hence, ${V}_{\rm eff}$ do not contribute to the scalar radiation.

Now, let us estimate how the localized energy around the I-ball/oscillon leaks out to $r\to \infty$.
The energy loss rate of the I-ball/oscillon is represented by
\begin{eqnarray}
    \frac{dE}{dt} =  4\pi r^2 T^{0r} \ ,
\end{eqnarray}
where  $T_{0r}$ denotes the Poynting vector given by,
\begin{eqnarray}
    T_{0r} = \q_0 \xi \q_r \xi \ .
\end{eqnarray}
By averaging over time, we obtain
\begin{eqnarray}
\label{eq:Tr0}
    4\pi r^2 \overline{T}_{r0}
       &=& \frac{-1}{2\pi} \left(
          \sum_{n\ge 2} \sum_{\{k\,|\, \omega_{nk}>m\}}
          \frac{g_{n+1}}{k!(n-k)!} \tilde\psi_n(\overline\omega_{nk}) \,
          \omega_{nk} \right) \nonumber \\
       &&\times  \left(
          \sum_{l\ge 2} \sum_{\{j\,|\, \omega_{lj}>m\}}
          \frac{g_{n+1}}{j!(l-j)!} \tilde\psi_l(\overline\omega_{nk}) \,
       \overline\omega_{nk} \right)
         \delta_{n-2k, l-2j}   \ ,
\end{eqnarray}
for $r\to \infty$.
By using this time-averaged Poynting vector, the decay rate of the I-ball/oscillon for a given energy $E$ and a charge $Q_0 = I_0$ is represented by
\begin{eqnarray}
    \Gamma = \frac{1}{E} | 4\pi r^2 \overline{T}_{r0} | \ ,
\end{eqnarray}
which is finite at $r\to \infty$.
By using $\Gamma$,  the lifetime of an I-ball/oscillon with an initial charge $Q_i = I_i$ is given by,
\begin{eqnarray}
\tau_I&=& \int_{Q_{i}}^{Q_{\rm cr}} \frac{\omega  dQ_0}{E(Q_0)\Gamma}\ ,
\end{eqnarray}
where $Q_{\rm cr}$ is the critical value of the charge below which no stable  I-ball/oscillon  exists (see the next subsection).

\subsection{Example}
\label{example}
\begin{figure}[tb]
    \begin{minipage}{.325\linewidth}
        \begin{center}
            \includegraphics[width=\linewidth]{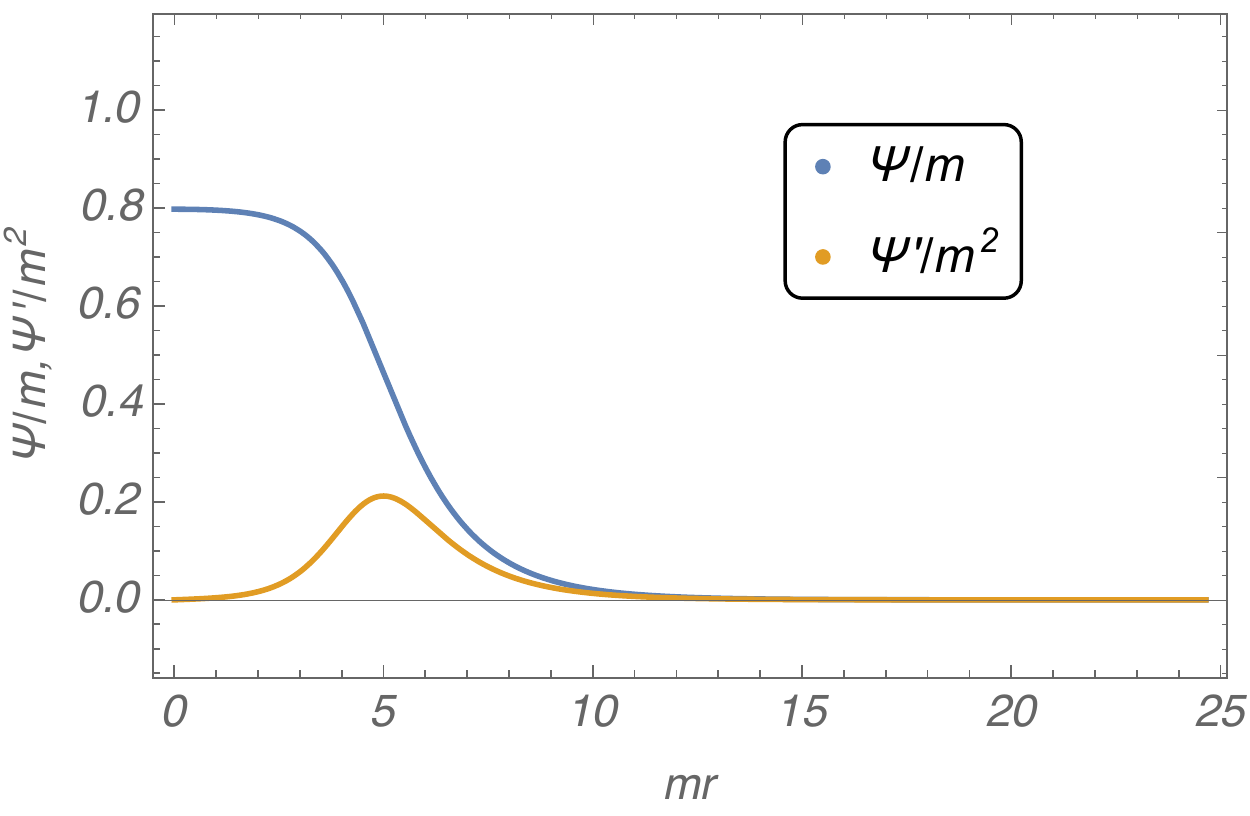}
        \end{center}
    \end{minipage}
    \begin{minipage}{.325\linewidth}
        \begin{center}
            \includegraphics[width=\linewidth]{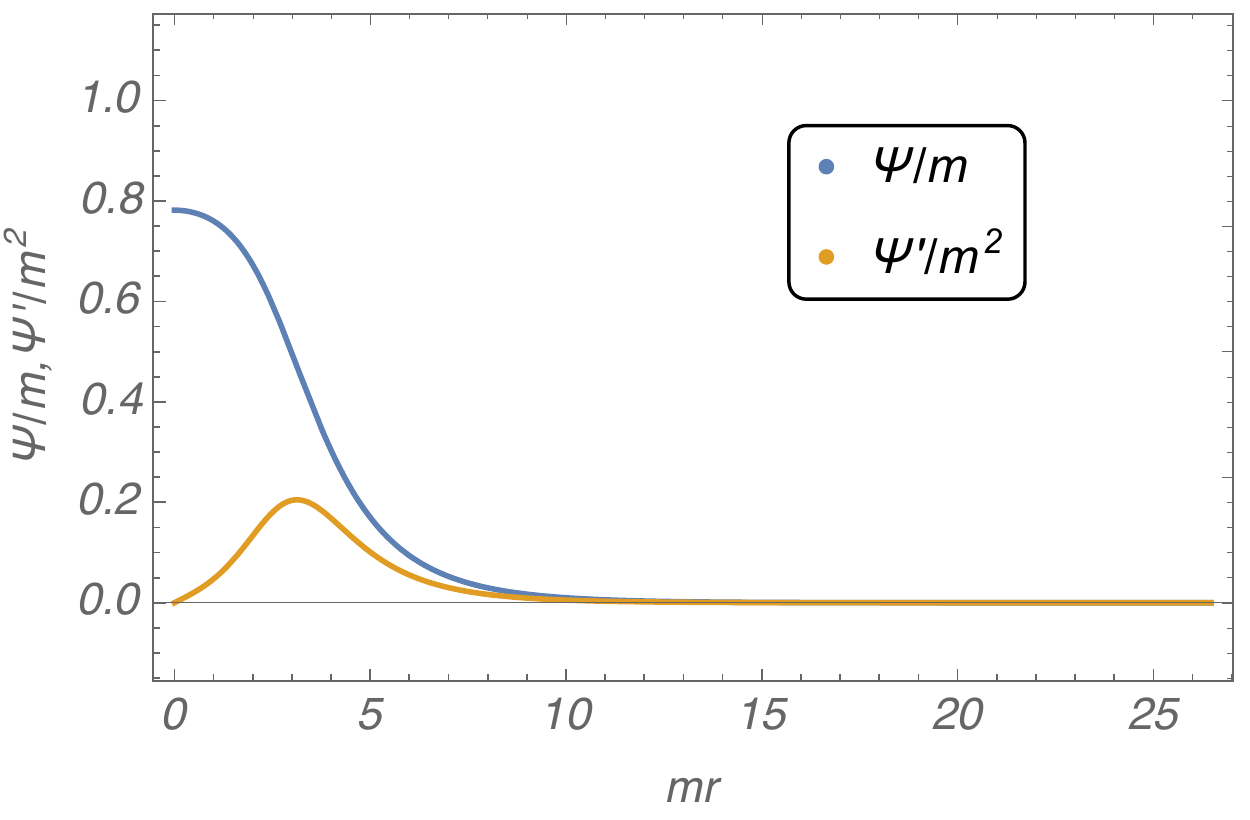}
        \end{center}
    \end{minipage}
    \begin{minipage}{.325\linewidth}
        \begin{center}
            \includegraphics[width=\linewidth]{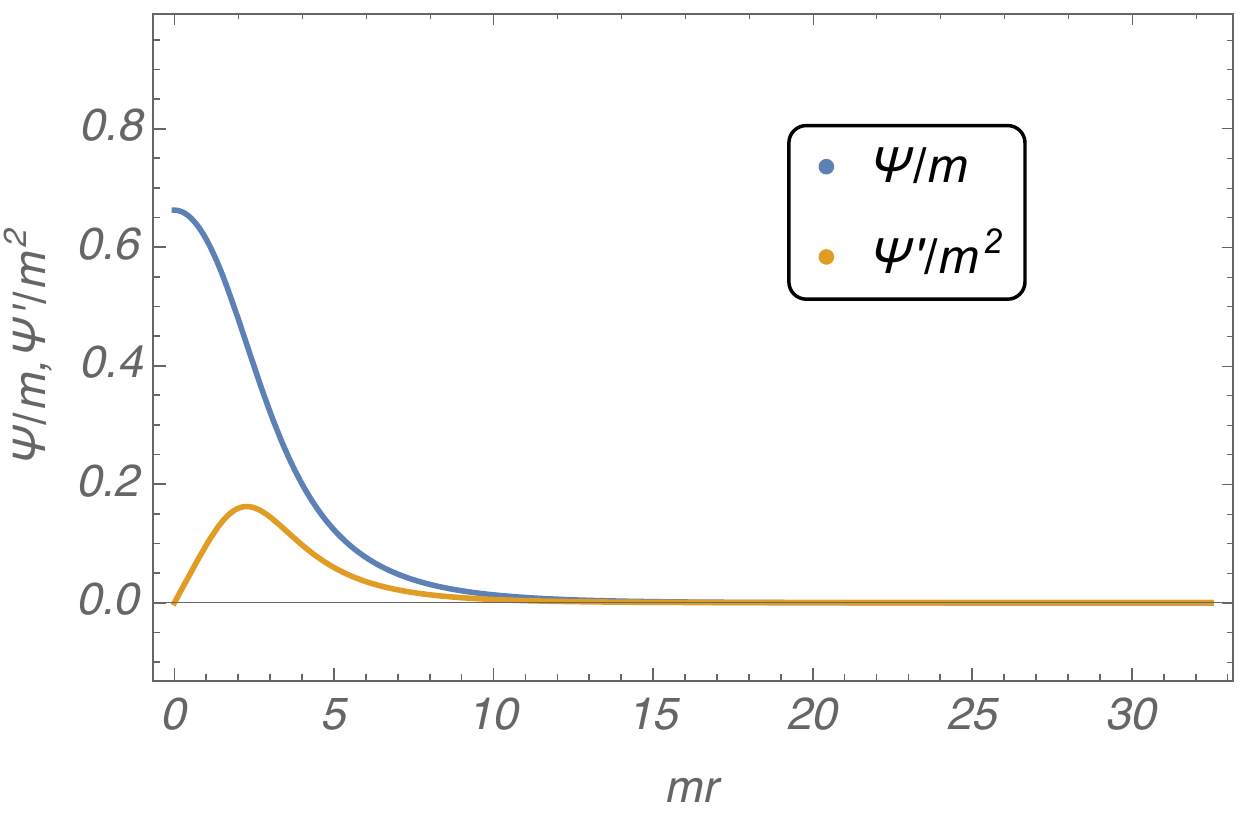}
        \end{center}
    \end{minipage}
    \caption{The I-ball/oscillon (Q-ball) solutions for a given $\omega$.
    In each panel, blue and yellow lines show $\psi/m$ and $d (\psi/m)/d(mr)$, respectively.
    The values of $\omega$ are $\omega = 0.85,\,0.90,\,0.95$ from left to right.
        }
    \label{fig:wf}
\end{figure}
Here we estimate the decay rate for a specific potential.
In the following, we consider
\begin{align}
    V (\phi) = \frac{g_4}{4!} \phi^4 + \frac{g_6}{6!} \phi^6  \ ,
    \label{eq:example_potential}
\end{align}
with $g_4 = -3!$ and $g_6 = 0.4 \times 5!\,m^{-2}$ to conform with the analysis in \cite{Mukaida:2016hwd}.
The scalar potential with these parameters satisfies the I-ball/oscillon (Q-ball) condition in Eq.\,(\ref{oscillon_condition}).
In Fig.\ref{fig:wf}, we show the I-ball/oscillon configuration for a given $\omega$.
It is seen that $\psi(r)$ is well described by the Gaussian profile for $\omega \gtrsim 0.9$.
The profile deviates from the Gaussian shape for a smaller $\omega$ (e.g. $\omega=0.85$).

\begin{figure}[t]
    \begin{minipage}{.42\linewidth}
        \begin{center}
            \includegraphics[width=\linewidth]{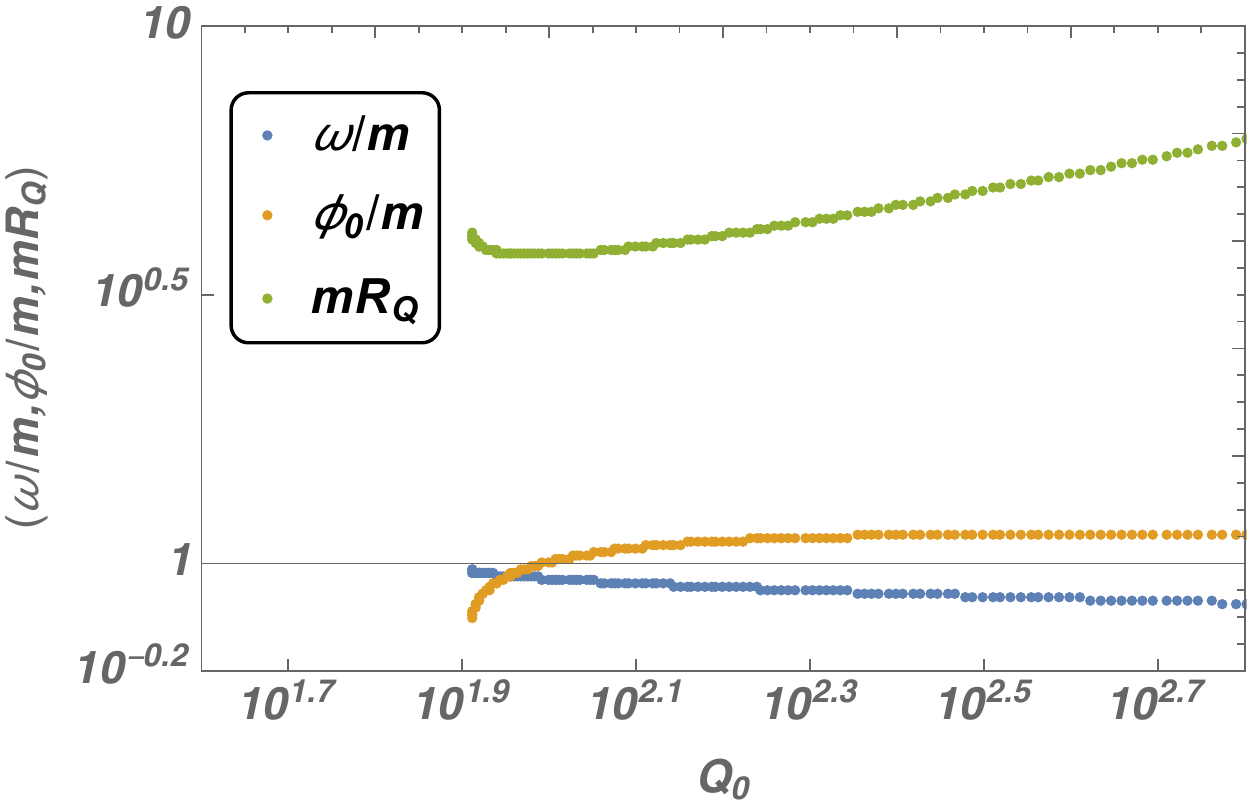}
        \end{center}
    \end{minipage}
    \begin{minipage}{.42\linewidth}
        \begin{center}
            \includegraphics[width=\linewidth]{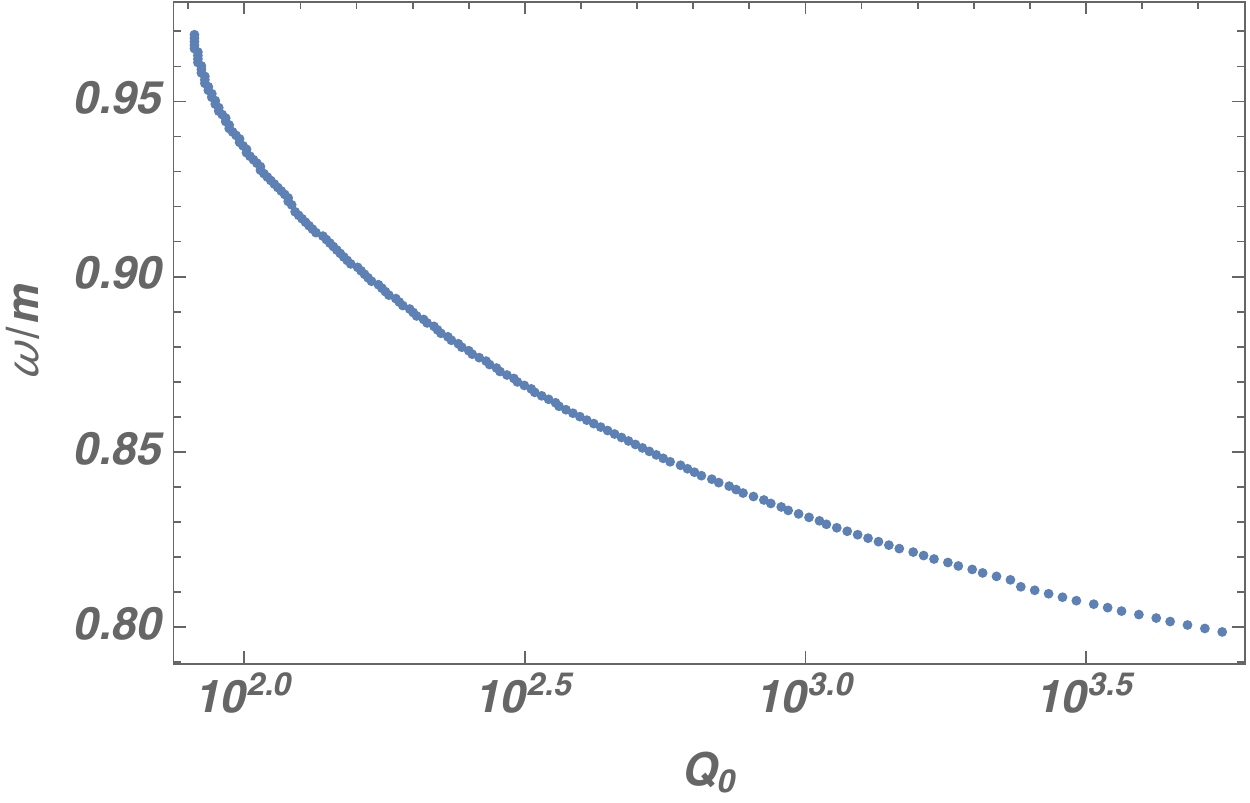}
        \end{center}
    \end{minipage}
       \caption{Left) The relations between $\omega$ (blue), $\phi_0 = \psi (r=0)$ (yellow), and $R_Q$ (green) and ${Q}_0 = I_0$.
     The normalization of the parameters $Q_0$ and $\psi_0$ are smaller than those in  \cite{Mukaida:2016hwd} by a factor of two.
     Right) The enlarged plot of $\omega$.
    }
    \label{fig:1}
\end{figure}

In Fig.\,\ref{fig:1}, we show $\omega$ (blue), $\psi_0 = \psi (r=0)$ (yellow), $R_Q$ (green) as functions of ${Q}_0 = I_0$, which reproduce Fig.\,1 in \cite{Mukaida:2016hwd}.%
\footnote{The normalizations of $\psi_0$ and $Q_0$ in this paper are different from those in Fig.~1 of  \cite{Mukaida:2016hwd} by a factor of two, respectively~\cite{Mukaida:PVC}.}
Here, $\psi_0$ and $R_Q$ are defined to fit the profile by a Gaussian profile,
\begin{eqnarray}
    \psi \sim \psi_0 \,\exp\left[-\frac{r^2}{R_{Q}^2}\right]\, .
\end{eqnarray}
In the figure, we show only the parameters for stable solutions, i.e. $d \omega/d Q_0< 0$~\cite{Mukaida:2016hwd}.%
\footnote{This condition corresponds to the condition for $E(Q_1 + Q_2) < E(Q_1) + E(Q_2)$ (see Eq.\,(\ref{eq:e-q})).}
There is no stable solution for the charges smaller than the critical value, $Q_{\rm cr} \simeq  10^{1.9}$.

\begin{figure}[t]
    \begin{minipage}{.32\linewidth}
        \begin{center}
            \includegraphics[width=\linewidth]{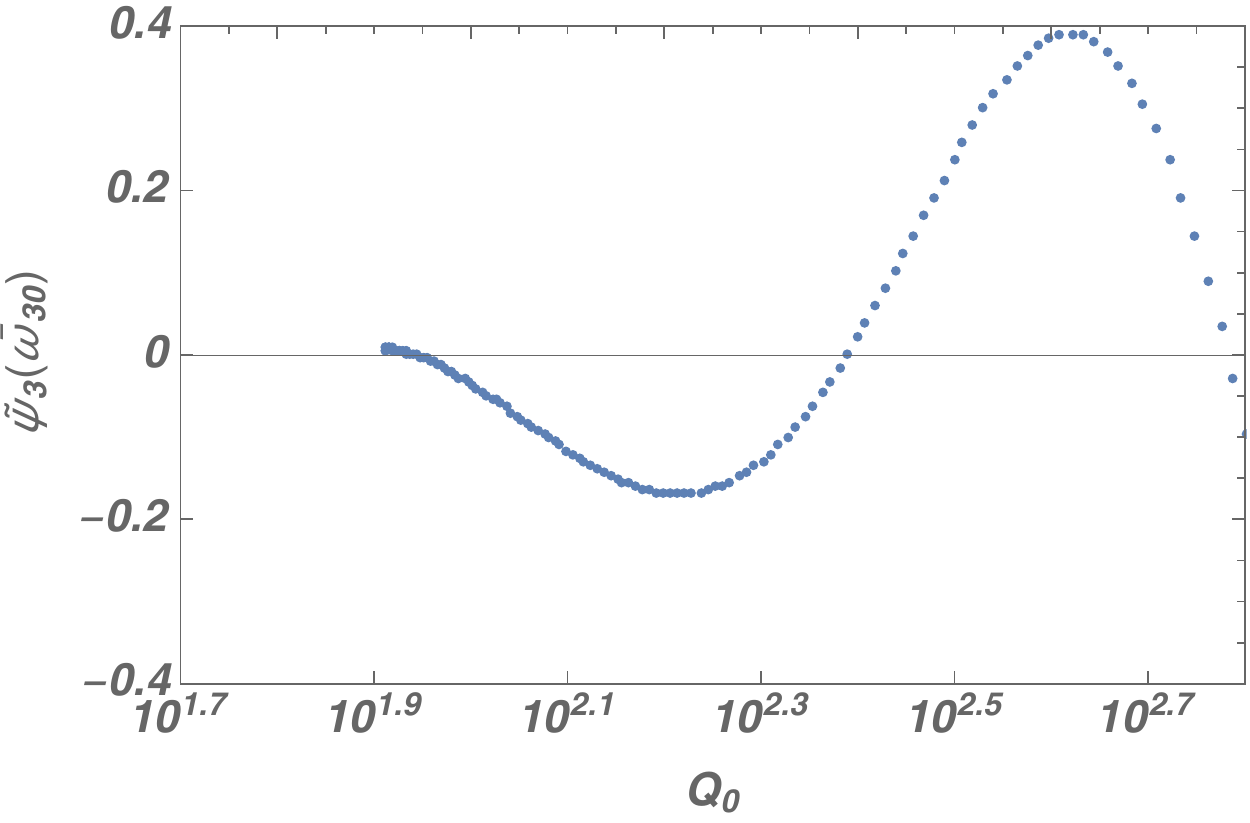}
        \end{center}
    \end{minipage}
    \begin{minipage}{.32\linewidth}
        \begin{center}
            \includegraphics[width=\linewidth]{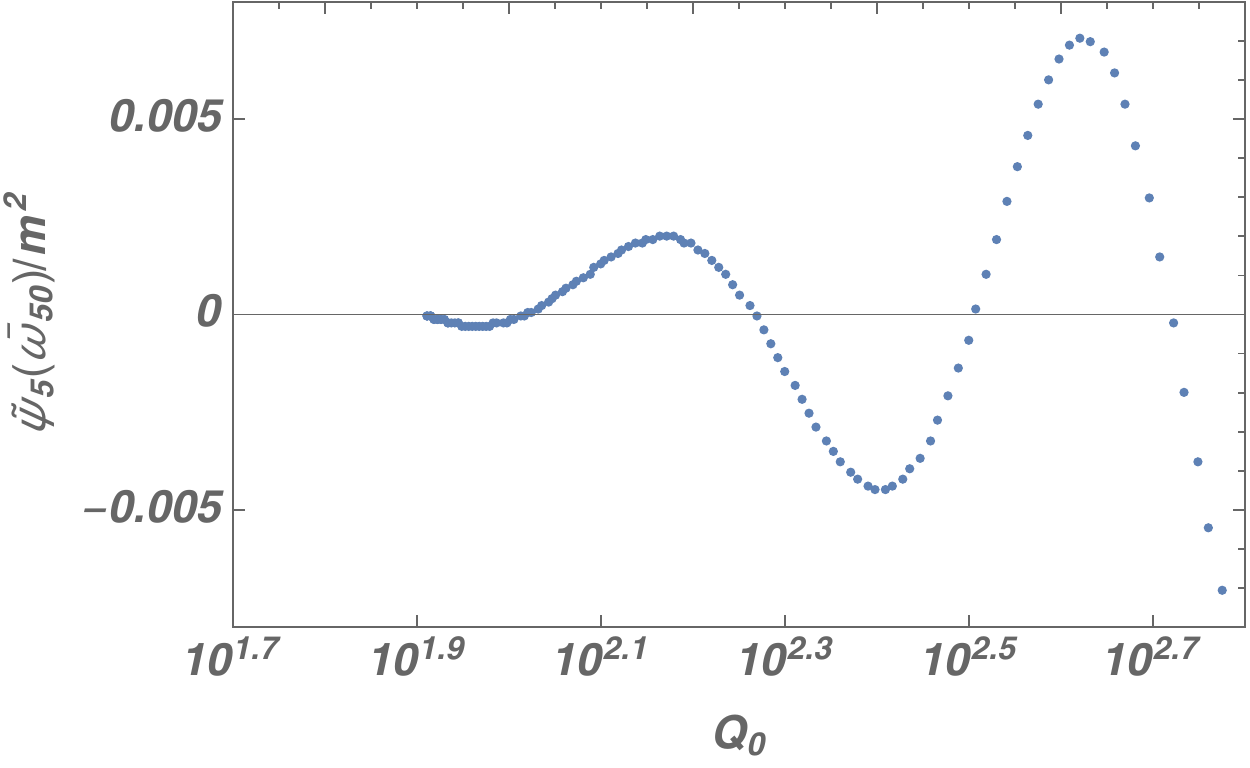}
        \end{center}
    \end{minipage}
    \begin{minipage}{.32\linewidth}
        \begin{center}
            \includegraphics[width=\linewidth]{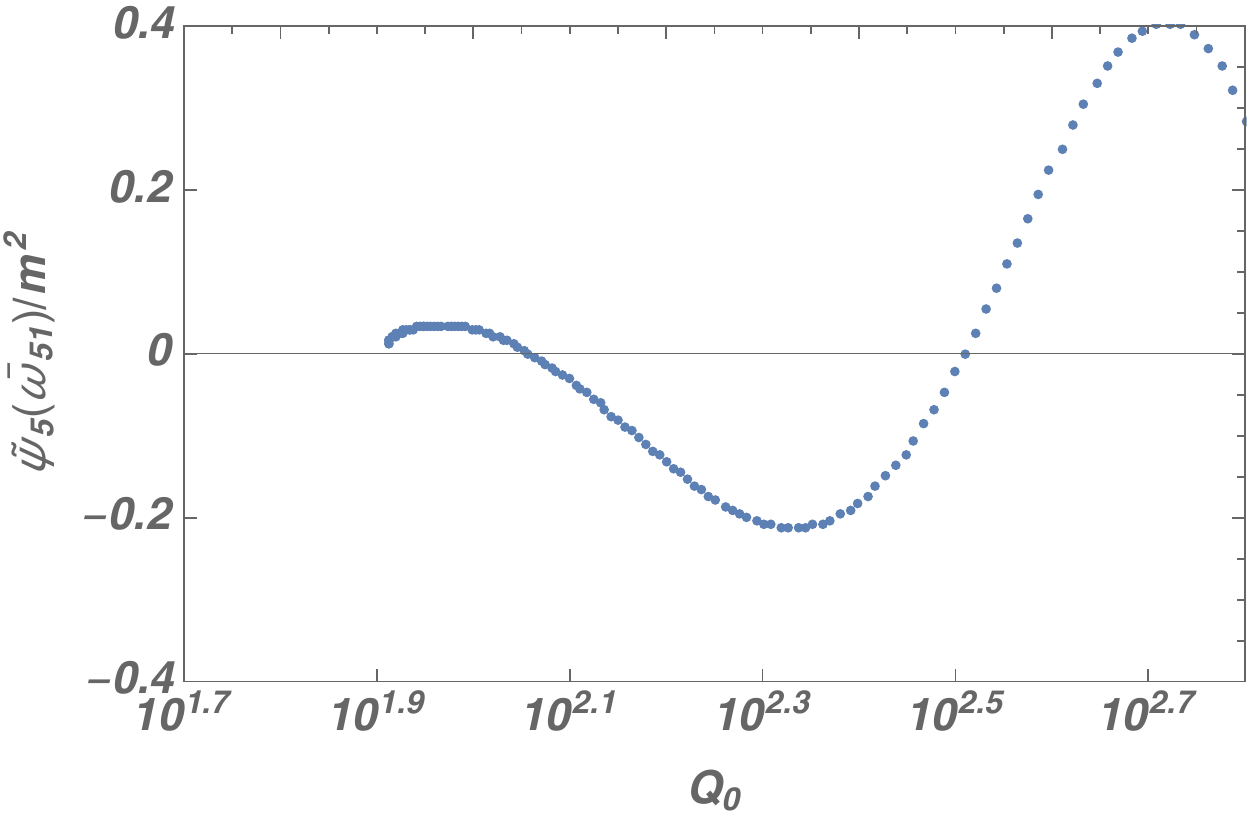}
        \end{center}
    \end{minipage}
    \caption{Plots of $\tilde\psi_{n}(\overline\omega_{nk})$ for given ${Q}_0$. We see that $\tilde\psi_{5}(\overline\omega_{50})$ is subdominant compared with $\tilde\psi_{3}(\overline\omega_{30})$ and
    $\tilde\psi_{5}(\overline\omega_{51})$.
    }
    \label{fig:psitil}
\end{figure}

We also plot $\tilde\psi_{n}(\overline\omega_{nk})$ for given ${Q}_0 = I_0$ in Fig.\,\ref{fig:psitil}.
The figure shows that $\tilde\psi_{5}(\overline\omega_{50})$ is subdominant compared with  $\tilde\psi_{3}(\overline\omega_{30})$ and
$\tilde\psi_{5}(\overline\omega_{51})$.
This can be understood as the emission of the mode of $\omega_{30} = \omega_{51} = 3\omega$ corresponds to the first excited state,
while that of $\omega_{50}$ to the second excited state.%
\footnote{The emission of the mode of $\omega$ is kinematically forbidden since $\omega < m$.
The mode of $2\omega$ is absent for the scalar potential with $g_{2n+1} = 0$.}

\begin{figure}[t]
    \begin{minipage}{.42\linewidth}
        \begin{center}
            \includegraphics[width=\linewidth]{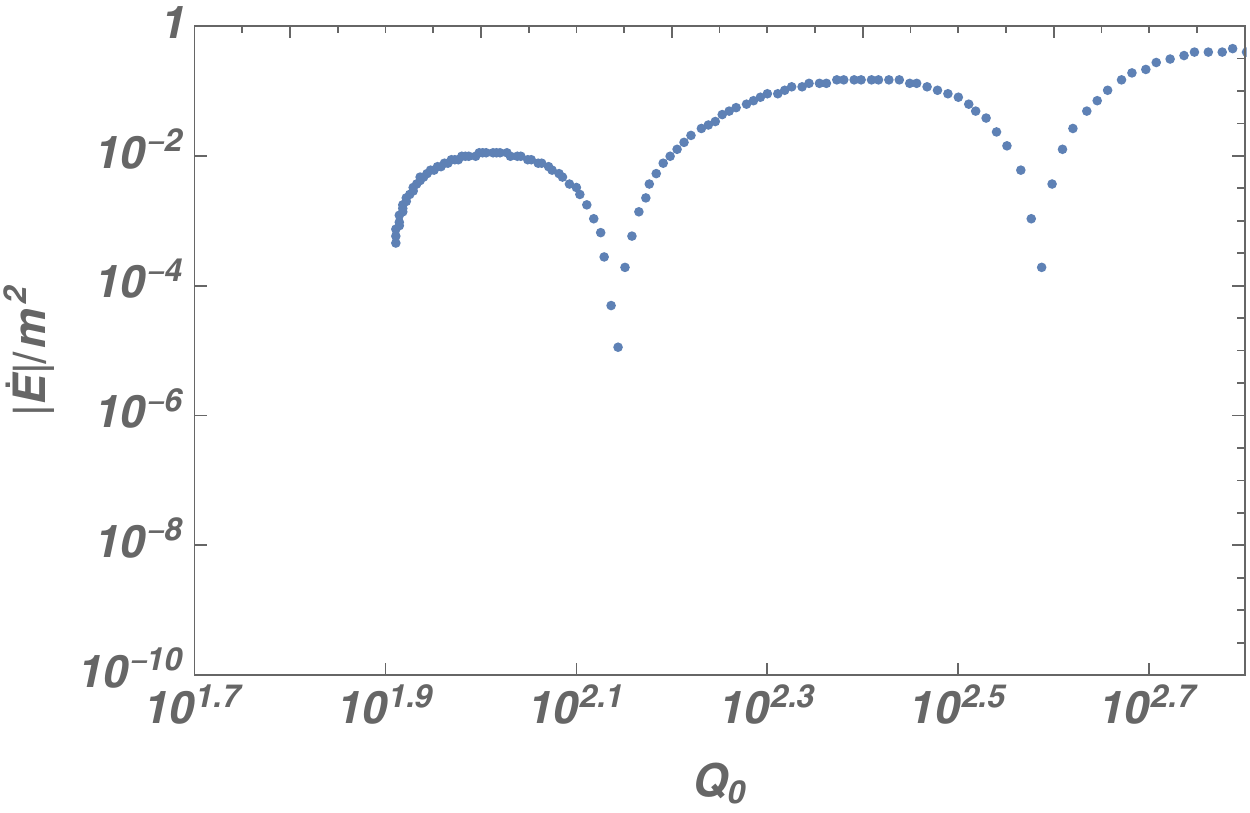}
        \end{center}
    \end{minipage}
    \begin{minipage}{.42\linewidth}
        \begin{center}
            \includegraphics[width=\linewidth]{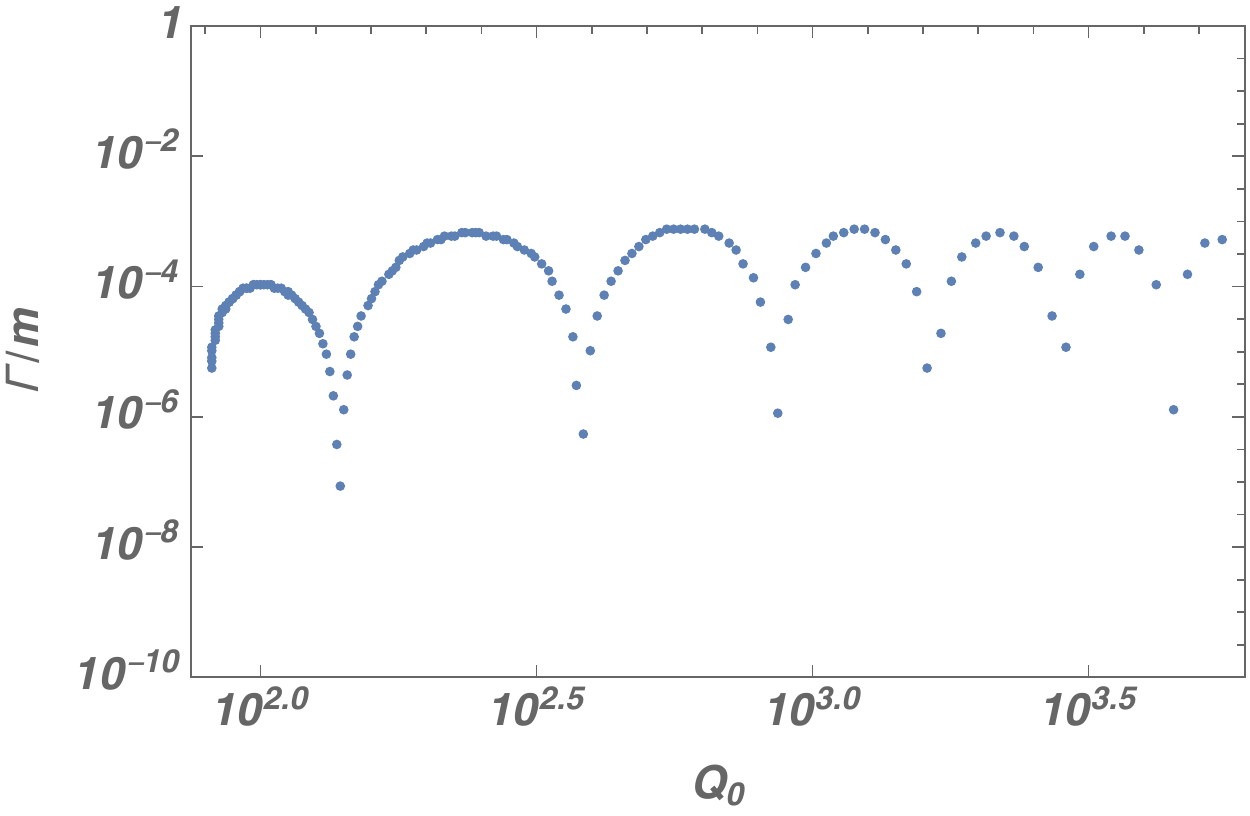}
        \end{center}
    \end{minipage}
    \caption{The time-derivative of the I-ball/oscillon energy (left) and the decay rate $\Gamma$ (right) for given ${Q}_0$.}
    \label{fig:Edot}
\end{figure}

Fig.\,\ref{fig:Edot} shows the absolute value of ${dE}/{dt}$ (left) and the decay rate $\Gamma$ (right) for given ${Q}_0$.
As the $\Gamma$ is dominated by the contributions form $\tilde\psi_{30}$ and $\tilde\psi_{51}$,
the position of the zeros of the decay rate  are determined by the zero points of $g_4\tilde\psi_{3}(\overline\omega_{30})/3! + g_6 \tilde\psi_{5}(\overline\omega_{51})/5!$,
though the decay rate is not exactly vanishing at the zero points  due to the contributions of other modes such as $\tilde\psi_{50}$.
\footnote{Here, we use
$\omega_{30} = \omega_{51}$ and $\overline\omega_{30} = \overline\omega_{51}$ in Eq.\,(\ref{eq:Tr0}).}
The I-ball/oscilon loses its energy gradually by emitting relativistic radiation with a give rate in the figure.
As there is no stable I-ball/oscillon solution below $Q_{\rm cr} \simeq 10^{1.9}$, the I-ball/oscillon
it rapidly decays once the charge reaches $Q_0 = I_0 = {Q}_{{\rm cr}}$ (see Sec.~\ref{sec:result}).%
\footnote{Although the decay rate is more less consistent with that of \cite{Mukaida:2016hwd} as a whole,
the zeros of the decay rate given are not well reproduced.
We validate our result by the classical lattice simulation in Sec.~\ref{sec:result}.}

\section{Validation of the analytic decay rate }
\label{sec:num_sim}

\subsection{Setup of Numerical Simulation}
To confirm the validity of the analytical calculation in the previous section, we perform a classical lattice simulation
of the time-evolution of a real scalar field $\phi$.
We calculate a relation between the I-ball/oscillon charge $Q_0 = I_0$ and the time derivative of the I-ball/oscillon energy $\dot{E}$.

In the simulation, units of energy and time are taken to be $m$ and $m^{-1}$, that is,
\begin{equation}
    \phi \rightarrow m\phi\ ,\ \
    t \rightarrow \frac{t}{m}\ ,\ \
    x \rightarrow \frac{x}{m}\ ,\ \dots\ {\rm etc}\ .
\end{equation}
We also assume that the configuration of $\phi$ is spherically symmetric in three spatial dimensions, so the equation of motion of $\phi$ is represented by
\begin{equation}
    \frac{d^2 \phi}{d t^2}
    = \frac{d^2 \phi}{d r^2} + \frac{2}{r}\frac{d \phi}{d r} - \frac{\partial V}{\partial \phi}\ .
    \label{eq:num_eom}
\end{equation}
The potential is the same as that adopted in \ref{example},
\begin{align}
    V (\phi) = \frac{g_4}{4!} \phi^4 + \frac{g_6}{6!} \phi^6 \ ,
\end{align}
where $g_4 = -3!$ and $ g_6 = 0.4 \times 5!$.
To avoid the divergence of the second term of the right-hand side of Eq.~(\ref{eq:num_eom}), we impose the following condition at the origin:
\begin{equation}
    \left. \frac{1}{r}\frac{d \phi}{d r} \right|_{r=0} = 0\,.
\end{equation}
At the boundary $r \to \infty$, we impose the absorbing boundary condition (see the appendix of the reference~\cite{Salmi:2012ta} for details).
Under this condition,  radiation of the real scalar field emitted from the I-ball/oscillon is absorbed at the boundary, so that we can calculate the dynamics of I-ball/oscillon correctly.

For the initial condition, we use the theoretical I-ball/oscillon profile for a given $\omega_{\rm ini}$ and
\begin{eqnarray}
    \dot{\phi}(t=0,r) &=& 0\ .
\end{eqnarray}
We choose $\omega_{\rm ini}$ properly to aquire the desired value of the I-ball/oscillon charge $Q_0$.
The other simulation parameters are shown in Table~\ref{Ta:simpara}.

{\renewcommand\arraystretch{1.1}
    \begin{table}[t]
        \centering
        \begin{tabular}{cc}
            \hline \hline
            $\omega_{\rm ini}$ & varying \tabularnewline
            $g_4$ & $-3!$ \tabularnewline
            $g_6$ & $0.4\times 5!$ \tabularnewline
            Box size $L$  & $64$ \tabularnewline
            Grid size $N$ & $1024$ \tabularnewline
            Initial time  & $0$ \tabularnewline
            Final time    & $1.0\times 10^5$ \tabularnewline
            Time step     & $1.0\times 10^{-2}$ \tabularnewline
            \hline \hline
        \end{tabular}
        \caption{
            Simulation parameters.
            $\omega_{\rm ini}$ is changed through simulations to set the appropriate initial value of the I-ball/oscillon charge $Q_0$.
        }
        \label{Ta:simpara}
    \end{table}
}

We develop our own classical lattice simulation code, in which the time evolution is calculated by the fourth-order symplectic integration scheme and  the spatial derivatives are by the fourth-order central difference scheme.
To check the correctness of the code, we have confirmed that the results do not significantly change when we set different simulation parameters (box size $L$, grid size $N$, time step $\Delta t$).

\subsection{Result}
\label{sec:result}
In numerical simulations, we cannot calculate $Q_0$ nor $I_0$ directly since $Q_0$ is defined by $\Psi$ while $I_0$ is defined
by an average over one period of the oscillation as in Eq.\,(\ref{eq:I}).
Instead, we approximate $Q_0=I_0$ by $Q$ defined by
\begin{eqnarray}
    Q &=& \frac{1}{T_{\rm ave}}
          \int ^t _{t-T_{\rm ave}} dt
          \int ^L _0 d^3x  \dot{\phi}^2, \\
      &=& \frac{1}{T_{\rm ave}}
          \int ^t _{t-T_{\rm ave}} dt
          \int ^L _0 dr 4\pi r^2 \dot{\phi}^2.
      \label{eq:numQ}
\end{eqnarray}
where $T_{\rm ave} = 100$ is the duration of the time average.
This value is much larger than $2\pi/\omega\simeq 10$, but much smaller than the typical time scale of the I-ball/oscillon decay $1/\Gamma \simeq 10^4$.
Thus $T_{\rm ave} = 100$ does not affect the results of our simulations.

We also take the time average to calculate the I-ball/oscillon energy
\begin{equation}
    E = \frac{1}{T_{\rm ave}} \int ^t _{t-T_{\rm ave}} dt
        \int ^L _0 dr 4\pi r^2
        \left[ \frac{1}{2}\dot{\phi}^2 + \frac{1}{2} (\nabla \phi)^2 +V \right],
    \label{eq:numE}
\end{equation}
and calculate $\Gamma = \dot{E}/E$ by the fourth-order central difference scheme.

\begin{figure}[t]
    \vspace{-0.3cm}
    \begin{center}
        \includegraphics[width=125mm]{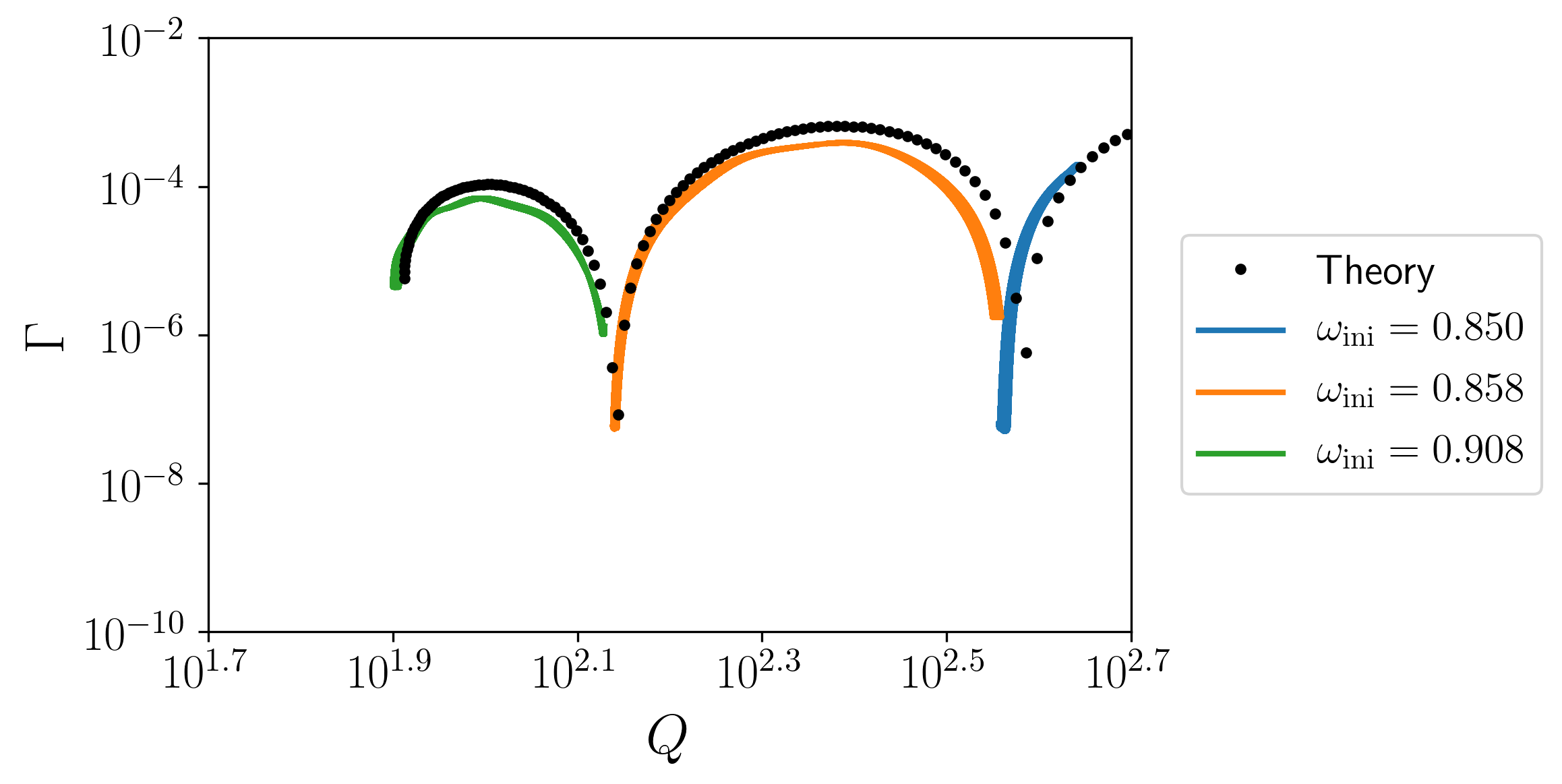}
    \end{center}
    \vspace{-0.5cm}
    \caption{
        Comparison between the results of our simulations and analytical calculation (Fig.~\ref{fig:Edot}).
        The definition of $Q$ and $E$ in the simulations are shown in Eq.\,(\ref{eq:numQ}) and Eq.\,(\ref{eq:numE}).
        Blue, orange, and green lines show the result of $\omega_{\rm ini} = 0.850$, $\omega_{\rm ini} = 0.858$, and $\omega_{\rm ini} = 0.908$ respectively
        and black dots show the analytical result.
        Because we set the final time of the numerical simulation as $t = 10^{5}$, the decay late smaller than $\sim 10^{-7}$ cannot be shown in this figure.
        We also cannot calculate the relation $Q\lesssim 10^{1.9}$
        because the field does not have the stable I-ball/oscillon solution in this range (see Fig.~\ref{fig:num_ex}), so we remove the data after the I-ball/oscillon decay for clarity. 
        The two results look slightly different in large charge ($Q \gtrsim 10^{2.5}$) because the approximation $\mu \ll m$ may not be appropriate as explained in \ref{example}. From this figure, we can find that the result of the analytical calculation is almost in agreement with the simulation results.
    }
    \label{fig:numerical_result}
\end{figure}

The results of the simulations are shown in Fig.~\ref{fig:numerical_result},
which are compared with our analytical calculation (see Fig.~\ref{fig:Edot}).
The figure shows that the analytical results are in good agreement with the results of the classical lattice simulation for ${Q}_0 \lesssim 10^{2.5}$.
On the other hand, for the I-ball/oscillon with a large charge ${Q}_0 \gtrsim 10^{2.5}$,
the lattice results deviate from the analytical results.
The deviation is partly because the approximation $\mu \ll m$ is no more valid for  ${Q}_0 \gtrsim 10^{2.5}$ (see~\ref{example}). Because we set the final time of the numerical simulation as $t = 10^{5}$, the decay late smaller than $\sim 10^{-7}$ cannot be shown in Fig.~\ref{fig:numerical_result}.

\begin{figure}[t]
    \vspace{-0.2cm}
    \begin{center}
        \includegraphics[width=95mm]{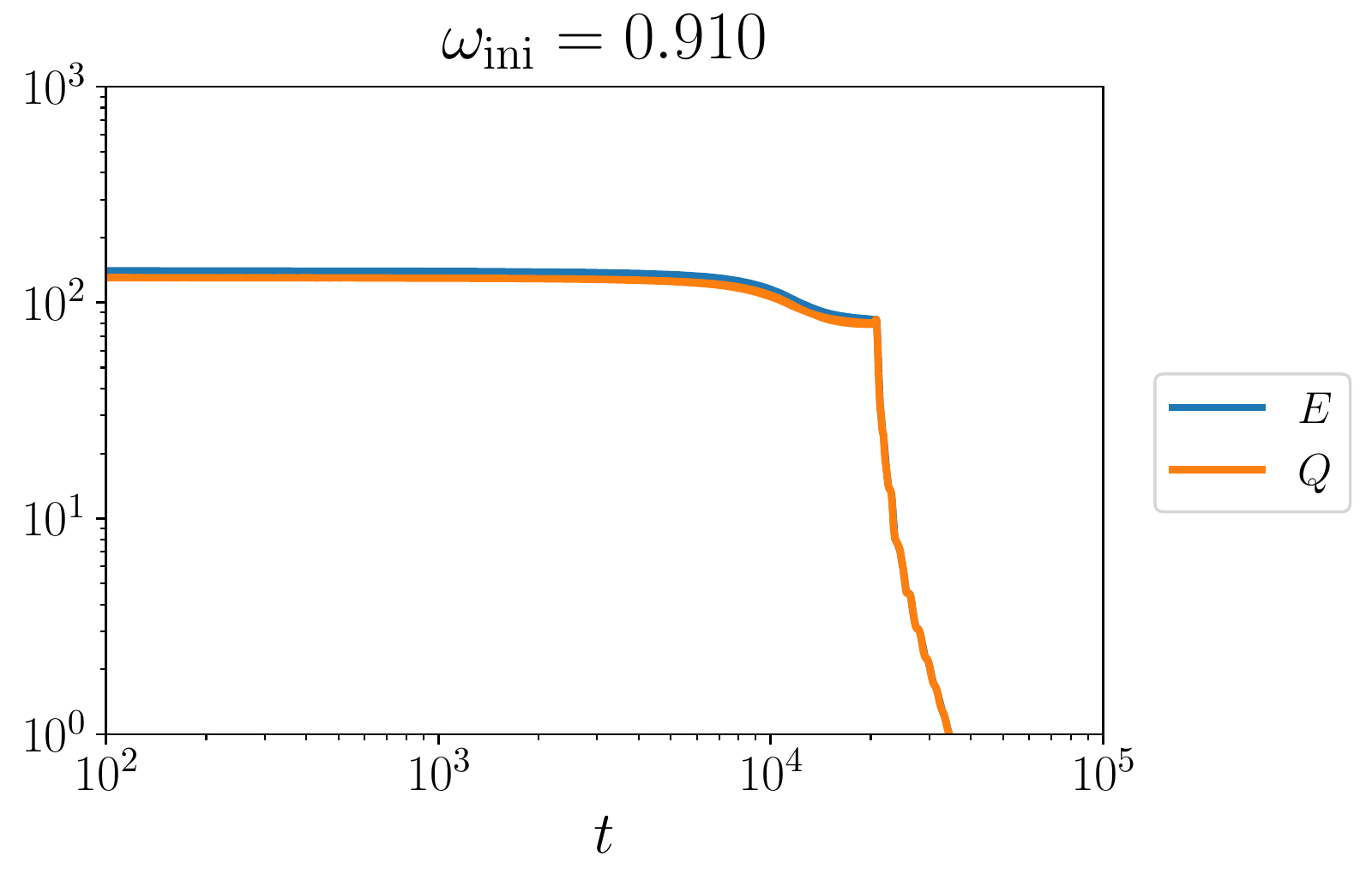}
    \end{center}
    \vspace{-0.5cm}
    \caption{
        One example of our simulations for $\omega_{\rm ini}=0.910$. Blue and orange lines show the I-ball/oscillon Energy $E$ and its charge $Q$.
        The figure shows that the I-ball/oscillon decays rapidly when its charge decreases down to $Q_{\rm cr}\simeq 10^{1.9}$ at $t\sim 10^4\times m^{-1}$.
    }
    \label{fig:num_ex}
\end{figure}

As we mentioned in the previous section, there is no stable  I-ball/oscillon solution for  $Q\lesssim 10^{1.9}\simeq 80$.
Accordingly, we expect that the I-ball/oscillon decays rapidly when its charge reaches $Q_{\rm cr}  \simeq 80$.
This situation is realized in the numerical simulation for $\omega_{\rm ini}=0.910$ as shown in Fig.~\ref{fig:num_ex}.
In this case, the I-ball/oscillon charge $Q$ becomes $10^{1.9}$ at $t\simeq 10^4$ and the I-ball/oscillon has completely decayed at $Q\simeq 80$ as expected.
This result is consistent with the analytical result Fig.~\ref{fig:Edot}.

\section{Conclusion}
\label{sec:con_dis}
In this paper, we have shown that the decay rate of the I-ball/oscillon within the classical field theory.
Our method applies to various scalar field models (potentials) that exhibit long-lived, spatially localized and time-dependent solutions.
Our analysis clarifies the decay process that it is just a leakage of the localized energy of the I-ball/oscillon via a classical emission of the relativistic modes of the scalar field.
From the point of view of the adiabatic charge, the decay process is caused by the deviation of the scalar potential from the quadratic one,
where the adiabatic invariant is not precisely conserved.
From the point of view of the  $U(1)$ charge, it corresponds to the $U(1)$ symmetry breaking due to the violation of the non-relativistic approximation by the emission of the relativistic modes.

To validate our analytical approach, we have performed a classical lattice simulation.
There, the classical relativistic field equation is solved by setting the initial condition of the real scalar field as an I-ball/oscillon configuration.
The results are in good agreement with the analytical result.
For ${Q}_0 \lesssim 10^{2.1}$, for example, the lifetime of I-ball/oscillon is $t \sim 10^4 m^{-1}$, which is expected from the estimation of the decay rate $\Gamma$ (See Fig.\,\ref{fig:Edot}).
The agreement between the analytical result and the numerical simulation shows that the leading order approximation in our analytical calculation is sufficient to obtain the decay rate of the I-ball/oscillon,
since the numerical calculation does not rely on the perturbative expansion of the solution around the I-ball/oscillon.

\begin{acknowledgments}
MI acknowledges useful communication with K.~Mukaida on the Q-ball interpretation of the I-ball/oscillon.
This work was supported by JSPS KAKENHI Grant Nos. 17H01131 (M.K.) and 17K05434 (M.K.), MEXT KAKENHI Grant Nos. 15H05889 (M.K.,M.I),
No. 16H03991(M.I.), No. 17H02878(M.I.), and No. 18H05542 (M.I.), and World Premier International Research Center Initiative (WPI Initiative), MEXT, Japan.
\end{acknowledgments}

\appendix

\section{Detail of Calculation of $\xi$}
\label{sec:calculation}

In this appendix, we show the details of the integration of $\xi$ used in \ref{Scalar_radiation}.
By using the retarded Green Function in Eq.\,(\ref{eq:retG}), the perturbation around the I-ball/oscillon solution is given by,
\begin{eqnarray}
    \xi(t,{\mathbf x})\!= \! \sum_{n\ge 1}\! \sum_{k = 0}^{n}
       \frac{\pi{}_nC_k}{2^n}
      \!\!\! \int\!  \frac{dp^0 d^3{\mathbf p}}{(2\pi)^4}
       \frac{e^{-ip^0 t + i {\mathbf p}\cdot{\mathbf x} }}
          {(p^0 +i\varepsilon )^2 \! -\! {\mathbf p}^2\! -\! m^2}
       \!\left(
         \! \delta(p^0 - \omega'_{nk})
          \!+\!\delta(p^0 + \omega'_{nk})
      \! \right)\!
       \tilde\rho_n(p)\ ,
\end{eqnarray}
where $\tilde\rho(-p) = \tilde\rho(p)$ and $\omega'_{nk} = (n-2k) \omega$.
We take the limit of $\varepsilon \to+ 0$ implicitly.
By integrating the delta functions of the source term by $p^0$ and
by integrating out the angular directions of ${\mathbf p}$,
we obtain
\begin{align}
    \xi(t,{\mathbf x})
        = &  \sum_{n\ge 1} \sum_{k = 0}^{n}
           \frac{\pi{}_nC_k}{2^n}
           \frac{i}{2(2\pi)^3}\int_{-\infty}^{\infty} dp
           \left(
               \frac{e^{-i\omega'_{nk} t  }}
               {(p-(\overline{\omega}_{nk}+i\varepsilon'))(p+(\overline{\omega}_{nk}+i\varepsilon'))}
              \right. \cr
           &  \quad\quad \quad\quad+
              \left.
                \frac{e^{i\omega'_{nk} t  }}
               {(p-(\overline{\omega}_{nk}-i\varepsilon'))(p+(\overline{\omega}_{nk}-i\varepsilon'))}
           \right)
         \times \frac{p}{r}\left(e^{i pr} - e^{-ipr}\right) \tilde\rho(p)\ ,
         \label{eq:intedxi}
\end{align}
where $\overline\omega_{nk} = \sqrt{\omega'^2_{nk} - m^2}$ and $\varepsilon' = (\omega'_{nk} / \overline\omega_{nk} ) \varepsilon$.
Here, we extend the integration region of $p$ from $(0,\infty)$ to $(-\infty,\infty)$ by using the fact that the integrand is an even function of $p$.

Now, let us perform integration with respect to $p$.
For $r > 0$, the integration of the terms proportional to $e^{ipr}$
can be performed by attaching an infinite arch to the integration contour in the complex plane of $p$ with Im$[p]>0$.
Thus, the integration is given by the residue of the poles in the Im$[p]>0$ region.
It should be noted that the poles with finite imaginary parts dump exponentially at $r\to \infty$.
In the following, we only keep the contributions with $\overline\omega_{nk}\in \mathbb R$, i.e. $|n - 2 k| > 1$, since
we are we are interested in the perturbation at $r\to \infty$.
As a result, the relevant poles are,
\begin{eqnarray}
 p &=& \overline\omega_{nk} + i \varepsilon' \ , \quad (\omega_{nk}' > 0)  \ ,\\
 p &=&  -\overline\omega_{nk} - i \varepsilon' \ , \quad (\omega_{nk}' < 0)  \ ,
\end{eqnarray}
for the first term in the first bracket of Eq.\,(\ref{eq:intedxi}),
and
 \begin{eqnarray}
 p &=&  - \overline\omega_{nk} + i \varepsilon' \ , \quad (\omega_{nk}' > 0)  \ ,\\
 p &=&  \overline\omega_{nk} - i \varepsilon' \ , \quad (\omega_{nk}' < 0)  \ ,
\end{eqnarray}
for the second term in the first bracket.
Similarly, the poles which contribute to the integration of the terms proportional to $e^{-ipr}$ are
\newpage
\begin{eqnarray}
 p &=& -\overline\omega_{nk} - i \varepsilon' \ , \quad (\omega_{nk}' > 0)  \ ,\\
 p &=&  \overline\omega_{nk} + i \varepsilon' \ , \quad (\omega_{nk}' < 0)  \ ,
\end{eqnarray}
for the first term in the first bracket of Eq.\,(\ref{eq:intedxi}),
and
 \begin{eqnarray}
 p &=&  \overline\omega_{nk} - i \varepsilon' \ , \quad (\omega_{nk}' > 0)  \ ,\\
 p &=&  -\overline\omega_{nk} + i \varepsilon' \ , \quad (\omega_{nk}' < 0)  \ ,
\end{eqnarray}

Altogether, the integration over $p$ leads to
\begin{eqnarray}
    \xi (t,{\mathbf x})
       &=& \sum_{n\ge 2} \sum_{\{k\,|\, m > \omega_{nk} \}}^n
           \frac{\pi}{2^n} {}_nC_k
           \frac{-1}{2\pi^2}    \frac{1}{r} \cos(\omega_{nk}t -     \overline{\omega}_{nk}r) \tilde\rho_n(\overline{\omega_{nk}})\ ,
     \end{eqnarray}
where we defined $\omega_{nk} = |(n-2k)\omega|$.
Here, we have neglected the contribution from $\rho_1$ in Eq.\,(\ref{eq:rho_1}), since $\omega_{1k} < m$ and Im$[\overline\omega_{1k}]\neq 0$ for $n = 1$.
By inserting Eq.\,(\ref{eq:wave}), we obtain
\begin{eqnarray}
    \xi (t,{\mathbf x})
       &=& \sum_{n\ge 1} \sum_{\{k\,|\, \omega_{nk} > m\}}^n
                 \frac{-1}{2\pi } \frac{g_{n+1}}{k!(n-k)!} \frac{1}{r} \cos(\omega_{nk}t -     \overline{\omega}_{nk}r) \tilde\psi_n(\overline{\omega_{nk}})\ , \\
               \tilde\psi_{n}(p)
      &=& 4\pi\int dr\,  \psi^n(r)  \frac{r \sin pr}{p}\ ,
     \end{eqnarray}
which gives Eq.\,(\ref{eq:ASsol}).

\bibliography{I-ball_decay}

\begin{thebibliography}{55}%
\makeatletter
\providecommand \@ifxundefined [1]{%
 \@ifx{#1\undefined}
}%
\providecommand \@ifnum [1]{%
 \ifnum #1\expandafter \@firstoftwo
 \else \expandafter \@secondoftwo
 \fi
}%
\providecommand \@ifx [1]{%
 \ifx #1\expandafter \@firstoftwo
 \else \expandafter \@secondoftwo
 \fi
}%
\providecommand \natexlab [1]{#1}%
\providecommand \enquote  [1]{``#1''}%
\providecommand \bibnamefont  [1]{#1}%
\providecommand \bibfnamefont [1]{#1}%
\providecommand \citenamefont [1]{#1}%
\providecommand \href@noop [0]{\@secondoftwo}%
\providecommand \href [0]{\begingroup \@sanitize@url \@href}%
\providecommand \@href[1]{\@@startlink{#1}\@@href}%
\providecommand \@@href[1]{\endgroup#1\@@endlink}%
\providecommand \@sanitize@url [0]{\catcode `\\12\catcode `\$12\catcode
  `\&12\catcode `\#12\catcode `\^12\catcode `\_12\catcode `\%12\relax}%
\providecommand \@@startlink[1]{}%
\providecommand \@@endlink[0]{}%
\providecommand \url  [0]{\begingroup\@sanitize@url \@url }%
\providecommand \@url [1]{\endgroup\@href {#1}{\urlprefix }}%
\providecommand \urlprefix  [0]{URL }%
\providecommand \Eprint [0]{\href }%
\providecommand \doibase [0]{http://dx.doi.org/}%
\providecommand \selectlanguage [0]{\@gobble}%
\providecommand \bibinfo  [0]{\@secondoftwo}%
\providecommand \bibfield  [0]{\@secondoftwo}%
\providecommand \translation [1]{[#1]}%
\providecommand \BibitemOpen [0]{}%
\providecommand \bibitemStop [0]{}%
\providecommand \bibitemNoStop [0]{.\EOS\space}%
\providecommand \EOS [0]{\spacefactor3000\relax}%
\providecommand \BibitemShut  [1]{\csname bibitem#1\endcsname}%
\let\auto@bib@innerbib\@empty
\bibitem [{\citenamefont {Guth}(1981)}]{Guth:1980zm}%
  \BibitemOpen
  \bibfield  {author} {\bibinfo {author} {\bibfnamefont {A.~H.}\ \bibnamefont
  {Guth}},\ }\href {\doibase 10.1103/PhysRevD.23.347} {\bibfield  {journal}
  {\bibinfo  {journal} {Phys. Rev.}\ }\textbf {\bibinfo {volume} {D23}},\
  \bibinfo {pages} {347} (\bibinfo {year} {1981})}\BibitemShut {NoStop}%
\bibitem [{\citenamefont {Linde}(1982)}]{Linde:1981mu}%
  \BibitemOpen
  \bibfield  {author} {\bibinfo {author} {\bibfnamefont {A.~D.}\ \bibnamefont
  {Linde}},\ }\href {\doibase 10.1016/0370-2693(82)91219-9} {\bibfield
  {journal} {\bibinfo  {journal} {Phys. Lett.}\ }\textbf {\bibinfo {volume}
  {108B}},\ \bibinfo {pages} {389} (\bibinfo {year} {1982})}\BibitemShut
  {NoStop}%
\bibitem [{\citenamefont {Albrecht}\ and\ \citenamefont
  {Steinhardt}(1982)}]{Albrecht:1982wi}%
  \BibitemOpen
  \bibfield  {author} {\bibinfo {author} {\bibfnamefont {A.}~\bibnamefont
  {Albrecht}}\ and\ \bibinfo {author} {\bibfnamefont {P.~J.}\ \bibnamefont
  {Steinhardt}},\ }\href {\doibase 10.1103/PhysRevLett.48.1220} {\bibfield
  {journal} {\bibinfo  {journal} {Phys. Rev. Lett.}\ }\textbf {\bibinfo
  {volume} {48}},\ \bibinfo {pages} {1220} (\bibinfo {year}
  {1982})}\BibitemShut {NoStop}%
\bibitem [{\citenamefont {Sato}(1981)}]{Sato:1980yn}%
  \BibitemOpen
  \bibfield  {author} {\bibinfo {author} {\bibfnamefont {K.}~\bibnamefont
  {Sato}},\ }\href@noop {} {\bibfield  {journal} {\bibinfo  {journal} {Mon.
  Not. Roy. Astron. Soc.}\ }\textbf {\bibinfo {volume} {195}},\ \bibinfo
  {pages} {467} (\bibinfo {year} {1981})}\BibitemShut {NoStop}%
\bibitem [{\citenamefont {Bogolyubsky}\ and\ \citenamefont
  {Makhankov}(1976)}]{Bogolyubsky:1976yu}%
  \BibitemOpen
  \bibfield  {author} {\bibinfo {author} {\bibfnamefont {I.~L.}\ \bibnamefont
  {Bogolyubsky}}\ and\ \bibinfo {author} {\bibfnamefont {V.~G.}\ \bibnamefont
  {Makhankov}},\ }\href@noop {} {\bibfield  {journal} {\bibinfo  {journal}
  {Pisma Zh. Eksp. Teor. Fiz.}\ }\textbf {\bibinfo {volume} {24}},\ \bibinfo
  {pages} {15} (\bibinfo {year} {1976})}\BibitemShut {NoStop}%
\bibitem [{\citenamefont {Gleiser}(1994)}]{Gleiser:1993pt}%
  \BibitemOpen
  \bibfield  {author} {\bibinfo {author} {\bibfnamefont {M.}~\bibnamefont
  {Gleiser}},\ }\href {\doibase 10.1103/PhysRevD.49.2978} {\bibfield  {journal}
  {\bibinfo  {journal} {Phys. Rev.}\ }\textbf {\bibinfo {volume} {D49}},\
  \bibinfo {pages} {2978} (\bibinfo {year} {1994})},\ \Eprint
  {http://arxiv.org/abs/hep-ph/9308279} {arXiv:hep-ph/9308279 [hep-ph]}
  \BibitemShut {NoStop}%
\bibitem [{\citenamefont {Copeland}\ \emph {et~al.}(1995)\citenamefont
  {Copeland}, \citenamefont {Gleiser},\ and\ \citenamefont
  {Muller}}]{Copeland:1995fq}%
  \BibitemOpen
  \bibfield  {author} {\bibinfo {author} {\bibfnamefont {E.~J.}\ \bibnamefont
  {Copeland}}, \bibinfo {author} {\bibfnamefont {M.}~\bibnamefont {Gleiser}}, \
  and\ \bibinfo {author} {\bibfnamefont {H.~R.}\ \bibnamefont {Muller}},\
  }\href {\doibase 10.1103/PhysRevD.52.1920} {\bibfield  {journal} {\bibinfo
  {journal} {Phys. Rev.}\ }\textbf {\bibinfo {volume} {D52}},\ \bibinfo {pages}
  {1920} (\bibinfo {year} {1995})},\ \Eprint
  {http://arxiv.org/abs/hep-ph/9503217} {arXiv:hep-ph/9503217 [hep-ph]}
  \BibitemShut {NoStop}%
\bibitem [{\citenamefont {Kasuya}\ \emph {et~al.}(2003)\citenamefont {Kasuya},
  \citenamefont {Kawasaki},\ and\ \citenamefont {Takahashi}}]{Kasuya:2002zs}%
  \BibitemOpen
  \bibfield  {author} {\bibinfo {author} {\bibfnamefont {S.}~\bibnamefont
  {Kasuya}}, \bibinfo {author} {\bibfnamefont {M.}~\bibnamefont {Kawasaki}}, \
  and\ \bibinfo {author} {\bibfnamefont {F.}~\bibnamefont {Takahashi}},\ }\href
  {\doibase 10.1016/S0370-2693(03)00344-7} {\bibfield  {journal} {\bibinfo
  {journal} {Phys. Lett.}\ }\textbf {\bibinfo {volume} {B559}},\ \bibinfo
  {pages} {99} (\bibinfo {year} {2003})},\ \Eprint
  {http://arxiv.org/abs/hep-ph/0209358} {arXiv:hep-ph/0209358 [hep-ph]}
  \BibitemShut {NoStop}%
\bibitem [{\citenamefont {Kawasaki}\ \emph {et~al.}(2015)\citenamefont
  {Kawasaki}, \citenamefont {Takahashi},\ and\ \citenamefont
  {Takeda}}]{Kawasaki:2015vga}%
  \BibitemOpen
  \bibfield  {author} {\bibinfo {author} {\bibfnamefont {M.}~\bibnamefont
  {Kawasaki}}, \bibinfo {author} {\bibfnamefont {F.}~\bibnamefont {Takahashi}},
  \ and\ \bibinfo {author} {\bibfnamefont {N.}~\bibnamefont {Takeda}},\ }\href
  {\doibase 10.1103/PhysRevD.92.105024} {\bibfield  {journal} {\bibinfo
  {journal} {Phys. Rev.}\ }\textbf {\bibinfo {volume} {D92}},\ \bibinfo {pages}
  {105024} (\bibinfo {year} {2015})},\ \Eprint
  {http://arxiv.org/abs/1508.01028} {arXiv:1508.01028 [hep-th]} \BibitemShut
  {NoStop}%
\bibitem [{\citenamefont {Zeldovich}\ \emph {et~al.}(1974)\citenamefont
  {Zeldovich}, \citenamefont {Kobzarev},\ and\ \citenamefont
  {Okun}}]{Zeldovich:1974uw}%
  \BibitemOpen
  \bibfield  {author} {\bibinfo {author} {\bibfnamefont {{\relax Ya}.~B.}\
  \bibnamefont {Zeldovich}}, \bibinfo {author} {\bibfnamefont {I.~{\relax
  Yu}.}\ \bibnamefont {Kobzarev}}, \ and\ \bibinfo {author} {\bibfnamefont
  {L.~B.}\ \bibnamefont {Okun}},\ }\href@noop {} {\bibfield  {journal}
  {\bibinfo  {journal} {Zh. Eksp. Teor. Fiz.}\ }\textbf {\bibinfo {volume}
  {67}},\ \bibinfo {pages} {3} (\bibinfo {year} {1974})},\ \bibinfo {note}
  {[Sov. Phys. JETP40,1(1974)]}\BibitemShut {NoStop}%
\bibitem [{\citenamefont {'t~Hooft}(1974)}]{tHooft:1974kcl}%
  \BibitemOpen
  \bibfield  {author} {\bibinfo {author} {\bibfnamefont {G.}~\bibnamefont
  {'t~Hooft}},\ }\href {\doibase 10.1016/0550-3213(74)90486-6} {\bibfield
  {journal} {\bibinfo  {journal} {Nucl. Phys.}\ }\textbf {\bibinfo {volume}
  {B79}},\ \bibinfo {pages} {276} (\bibinfo {year} {1974})},\ \bibinfo {note}
  {[,291(1974)]}\BibitemShut {NoStop}%
\bibitem [{\citenamefont {Polyakov}(1974)}]{Polyakov:1974ek}%
  \BibitemOpen
  \bibfield  {author} {\bibinfo {author} {\bibfnamefont {A.~M.}\ \bibnamefont
  {Polyakov}},\ }\href@noop {} {\bibfield  {journal} {\bibinfo  {journal} {JETP
  Lett.}\ }\textbf {\bibinfo {volume} {20}},\ \bibinfo {pages} {194} (\bibinfo
  {year} {1974})},\ \bibinfo {note} {[,300(1974)]}\BibitemShut {NoStop}%
\bibitem [{\citenamefont {Kibble}(1976)}]{Kibble:1976sj}%
  \BibitemOpen
  \bibfield  {author} {\bibinfo {author} {\bibfnamefont {T.~W.~B.}\
  \bibnamefont {Kibble}},\ }\href {\doibase 10.1088/0305-4470/9/8/029}
  {\bibfield  {journal} {\bibinfo  {journal} {J. Phys.}\ }\textbf {\bibinfo
  {volume} {A9}},\ \bibinfo {pages} {1387} (\bibinfo {year}
  {1976})}\BibitemShut {NoStop}%
\bibitem [{\citenamefont {Coleman}(1985)}]{Coleman:1985ki}%
  \BibitemOpen
  \bibfield  {author} {\bibinfo {author} {\bibfnamefont {S.~R.}\ \bibnamefont
  {Coleman}},\ }\href {\doibase 10.1016/0550-3213(85)90286-X,
  10.1016/0550-3213(86)90520-1} {\bibfield  {journal} {\bibinfo  {journal}
  {Nucl. Phys.}\ }\textbf {\bibinfo {volume} {B262}},\ \bibinfo {pages} {263}
  (\bibinfo {year} {1985})},\ \bibinfo {note} {[Erratum: Nucl.
  Phys.B269,744(1986)]}\BibitemShut {NoStop}%
\bibitem [{\citenamefont {Kusenko}\ and\ \citenamefont
  {Shaposhnikov}(1998)}]{Kusenko:1997si}%
  \BibitemOpen
  \bibfield  {author} {\bibinfo {author} {\bibfnamefont {A.}~\bibnamefont
  {Kusenko}}\ and\ \bibinfo {author} {\bibfnamefont {M.~E.}\ \bibnamefont
  {Shaposhnikov}},\ }\href {\doibase 10.1016/S0370-2693(97)01375-0} {\bibfield
  {journal} {\bibinfo  {journal} {Phys. Lett.}\ }\textbf {\bibinfo {volume}
  {B418}},\ \bibinfo {pages} {46} (\bibinfo {year} {1998})},\ \Eprint
  {http://arxiv.org/abs/hep-ph/9709492} {arXiv:hep-ph/9709492 [hep-ph]}
  \BibitemShut {NoStop}%
\bibitem [{\citenamefont {Enqvist}\ and\ \citenamefont
  {McDonald}(1998)}]{Enqvist:1997si}%
  \BibitemOpen
  \bibfield  {author} {\bibinfo {author} {\bibfnamefont {K.}~\bibnamefont
  {Enqvist}}\ and\ \bibinfo {author} {\bibfnamefont {J.}~\bibnamefont
  {McDonald}},\ }\href {\doibase 10.1016/S0370-2693(98)00271-8} {\bibfield
  {journal} {\bibinfo  {journal} {Phys. Lett.}\ }\textbf {\bibinfo {volume}
  {B425}},\ \bibinfo {pages} {309} (\bibinfo {year} {1998})},\ \Eprint
  {http://arxiv.org/abs/hep-ph/9711514} {arXiv:hep-ph/9711514 [hep-ph]}
  \BibitemShut {NoStop}%
\bibitem [{\citenamefont {Enqvist}\ and\ \citenamefont
  {McDonald}(1999)}]{Enqvist:1998en}%
  \BibitemOpen
  \bibfield  {author} {\bibinfo {author} {\bibfnamefont {K.}~\bibnamefont
  {Enqvist}}\ and\ \bibinfo {author} {\bibfnamefont {J.}~\bibnamefont
  {McDonald}},\ }\href {\doibase 10.1016/S0550-3213(98)00695-6} {\bibfield
  {journal} {\bibinfo  {journal} {Nucl. Phys.}\ }\textbf {\bibinfo {volume}
  {B538}},\ \bibinfo {pages} {321} (\bibinfo {year} {1999})},\ \Eprint
  {http://arxiv.org/abs/hep-ph/9803380} {arXiv:hep-ph/9803380 [hep-ph]}
  \BibitemShut {NoStop}%
\bibitem [{\citenamefont {Kasuya}\ and\ \citenamefont
  {Kawasaki}(2000{\natexlab{a}})}]{Kasuya:1999wu}%
  \BibitemOpen
  \bibfield  {author} {\bibinfo {author} {\bibfnamefont {S.}~\bibnamefont
  {Kasuya}}\ and\ \bibinfo {author} {\bibfnamefont {M.}~\bibnamefont
  {Kawasaki}},\ }\href {\doibase 10.1103/PhysRevD.61.041301} {\bibfield
  {journal} {\bibinfo  {journal} {Phys. Rev.}\ }\textbf {\bibinfo {volume}
  {D61}},\ \bibinfo {pages} {041301} (\bibinfo {year} {2000}{\natexlab{a}})},\
  \Eprint {http://arxiv.org/abs/hep-ph/9909509} {arXiv:hep-ph/9909509 [hep-ph]}
  \BibitemShut {NoStop}%
\bibitem [{\citenamefont {Kasuya}\ and\ \citenamefont
  {Kawasaki}(2000{\natexlab{b}})}]{Kasuya:2000wx}%
  \BibitemOpen
  \bibfield  {author} {\bibinfo {author} {\bibfnamefont {S.}~\bibnamefont
  {Kasuya}}\ and\ \bibinfo {author} {\bibfnamefont {M.}~\bibnamefont
  {Kawasaki}},\ }\href {\doibase 10.1103/PhysRevD.62.023512} {\bibfield
  {journal} {\bibinfo  {journal} {Phys. Rev.}\ }\textbf {\bibinfo {volume}
  {D62}},\ \bibinfo {pages} {023512} (\bibinfo {year} {2000}{\natexlab{b}})},\
  \Eprint {http://arxiv.org/abs/hep-ph/0002285} {arXiv:hep-ph/0002285 [hep-ph]}
  \BibitemShut {NoStop}%
\bibitem [{\citenamefont {Mukaida}\ and\ \citenamefont
  {Takimoto}(2014)}]{Mukaida:2014oza}%
  \BibitemOpen
  \bibfield  {author} {\bibinfo {author} {\bibfnamefont {K.}~\bibnamefont
  {Mukaida}}\ and\ \bibinfo {author} {\bibfnamefont {M.}~\bibnamefont
  {Takimoto}},\ }\href {\doibase 10.1088/1475-7516/2014/08/051} {\bibfield
  {journal} {\bibinfo  {journal} {JCAP}\ }\textbf {\bibinfo {volume} {1408}},\
  \bibinfo {pages} {051} (\bibinfo {year} {2014})},\ \Eprint
  {http://arxiv.org/abs/1405.3233} {arXiv:1405.3233 [hep-ph]} \BibitemShut
  {NoStop}%
\bibitem [{\citenamefont {McDonald}(2002)}]{McDonald:2001iv}%
  \BibitemOpen
  \bibfield  {author} {\bibinfo {author} {\bibfnamefont {J.}~\bibnamefont
  {McDonald}},\ }\href {\doibase 10.1103/PhysRevD.66.043525} {\bibfield
  {journal} {\bibinfo  {journal} {Phys. Rev.}\ }\textbf {\bibinfo {volume}
  {D66}},\ \bibinfo {pages} {043525} (\bibinfo {year} {2002})},\ \Eprint
  {http://arxiv.org/abs/hep-ph/0105235} {arXiv:hep-ph/0105235 [hep-ph]}
  \BibitemShut {NoStop}%
\bibitem [{\citenamefont {Amin}\ and\ \citenamefont
  {Shirokoff}(2010)}]{Amin:2010jq}%
  \BibitemOpen
  \bibfield  {author} {\bibinfo {author} {\bibfnamefont {M.~A.}\ \bibnamefont
  {Amin}}\ and\ \bibinfo {author} {\bibfnamefont {D.}~\bibnamefont
  {Shirokoff}},\ }\href {\doibase 10.1103/PhysRevD.81.085045} {\bibfield
  {journal} {\bibinfo  {journal} {Phys. Rev.}\ }\textbf {\bibinfo {volume}
  {D81}},\ \bibinfo {pages} {085045} (\bibinfo {year} {2010})},\ \Eprint
  {http://arxiv.org/abs/1002.3380} {arXiv:1002.3380 [astro-ph.CO]} \BibitemShut
  {NoStop}%
\bibitem [{\citenamefont {Amin}\ \emph {et~al.}(2012)\citenamefont {Amin},
  \citenamefont {Easther}, \citenamefont {Finkel}, \citenamefont {Flauger},\
  and\ \citenamefont {Hertzberg}}]{Amin:2011hj}%
  \BibitemOpen
  \bibfield  {author} {\bibinfo {author} {\bibfnamefont {M.~A.}\ \bibnamefont
  {Amin}}, \bibinfo {author} {\bibfnamefont {R.}~\bibnamefont {Easther}},
  \bibinfo {author} {\bibfnamefont {H.}~\bibnamefont {Finkel}}, \bibinfo
  {author} {\bibfnamefont {R.}~\bibnamefont {Flauger}}, \ and\ \bibinfo
  {author} {\bibfnamefont {M.~P.}\ \bibnamefont {Hertzberg}},\ }\href {\doibase
  10.1103/PhysRevLett.108.241302} {\bibfield  {journal} {\bibinfo  {journal}
  {Phys. Rev. Lett.}\ }\textbf {\bibinfo {volume} {108}},\ \bibinfo {pages}
  {241302} (\bibinfo {year} {2012})},\ \Eprint {http://arxiv.org/abs/1106.3335}
  {arXiv:1106.3335 [astro-ph.CO]} \BibitemShut {NoStop}%
\bibitem [{\citenamefont {Amin}(2013)}]{Amin:2013ika}%
  \BibitemOpen
  \bibfield  {author} {\bibinfo {author} {\bibfnamefont {M.~A.}\ \bibnamefont
  {Amin}},\ }\href {\doibase 10.1103/PhysRevD.87.123505} {\bibfield  {journal}
  {\bibinfo  {journal} {Phys. Rev.}\ }\textbf {\bibinfo {volume} {D87}},\
  \bibinfo {pages} {123505} (\bibinfo {year} {2013})},\ \Eprint
  {http://arxiv.org/abs/1303.1102} {arXiv:1303.1102 [astro-ph.CO]} \BibitemShut
  {NoStop}%
\bibitem [{\citenamefont {Takeda}\ and\ \citenamefont
  {Watanabe}(2014)}]{Takeda:2014qma}%
  \BibitemOpen
  \bibfield  {author} {\bibinfo {author} {\bibfnamefont {N.}~\bibnamefont
  {Takeda}}\ and\ \bibinfo {author} {\bibfnamefont {Y.}~\bibnamefont
  {Watanabe}},\ }\href {\doibase 10.1103/PhysRevD.90.023519} {\bibfield
  {journal} {\bibinfo  {journal} {Phys. Rev.}\ }\textbf {\bibinfo {volume}
  {D90}},\ \bibinfo {pages} {023519} (\bibinfo {year} {2014})},\ \Eprint
  {http://arxiv.org/abs/1405.3830} {arXiv:1405.3830 [astro-ph.CO]} \BibitemShut
  {NoStop}%
\bibitem [{\citenamefont {Lozanov}\ and\ \citenamefont
  {Amin}(2017)}]{Lozanov:2016hid}%
  \BibitemOpen
  \bibfield  {author} {\bibinfo {author} {\bibfnamefont {K.~D.}\ \bibnamefont
  {Lozanov}}\ and\ \bibinfo {author} {\bibfnamefont {M.~A.}\ \bibnamefont
  {Amin}},\ }\href {\doibase 10.1103/PhysRevLett.119.061301} {\bibfield
  {journal} {\bibinfo  {journal} {Phys. Rev. Lett.}\ }\textbf {\bibinfo
  {volume} {119}},\ \bibinfo {pages} {061301} (\bibinfo {year} {2017})},\
  \Eprint {http://arxiv.org/abs/1608.01213} {arXiv:1608.01213 [astro-ph.CO]}
  \BibitemShut {NoStop}%
\bibitem [{\citenamefont {Hasegawa}\ and\ \citenamefont
  {Hong}(2018)}]{Hasegawa:2017iay}%
  \BibitemOpen
  \bibfield  {author} {\bibinfo {author} {\bibfnamefont {F.}~\bibnamefont
  {Hasegawa}}\ and\ \bibinfo {author} {\bibfnamefont {J.-P.}\ \bibnamefont
  {Hong}},\ }\href {\doibase 10.1103/PhysRevD.97.083514} {\bibfield  {journal}
  {\bibinfo  {journal} {Phys. Rev.}\ }\textbf {\bibinfo {volume} {D97}},\
  \bibinfo {pages} {083514} (\bibinfo {year} {2018})},\ \Eprint
  {http://arxiv.org/abs/1710.07487} {arXiv:1710.07487 [astro-ph.CO]}
  \BibitemShut {NoStop}%
\bibitem [{\citenamefont {Antusch}\ \emph {et~al.}(2018)\citenamefont
  {Antusch}, \citenamefont {Cefala}, \citenamefont {Krippendorf}, \citenamefont
  {Muia}, \citenamefont {Orani},\ and\ \citenamefont
  {Quevedo}}]{Antusch:2017flz}%
  \BibitemOpen
  \bibfield  {author} {\bibinfo {author} {\bibfnamefont {S.}~\bibnamefont
  {Antusch}}, \bibinfo {author} {\bibfnamefont {F.}~\bibnamefont {Cefala}},
  \bibinfo {author} {\bibfnamefont {S.}~\bibnamefont {Krippendorf}}, \bibinfo
  {author} {\bibfnamefont {F.}~\bibnamefont {Muia}}, \bibinfo {author}
  {\bibfnamefont {S.}~\bibnamefont {Orani}}, \ and\ \bibinfo {author}
  {\bibfnamefont {F.}~\bibnamefont {Quevedo}},\ }\href {\doibase
  10.1007/JHEP01(2018)083} {\bibfield  {journal} {\bibinfo  {journal} {JHEP}\
  }\textbf {\bibinfo {volume} {01}},\ \bibinfo {pages} {083} (\bibinfo {year}
  {2018})},\ \Eprint {http://arxiv.org/abs/1708.08922} {arXiv:1708.08922
  [hep-th]} \BibitemShut {NoStop}%
\bibitem [{\citenamefont {Hong}\ \emph {et~al.}(2018)\citenamefont {Hong},
  \citenamefont {Kawasaki},\ and\ \citenamefont {Yamazaki}}]{Hong:2017ooe}%
  \BibitemOpen
  \bibfield  {author} {\bibinfo {author} {\bibfnamefont {J.-P.}\ \bibnamefont
  {Hong}}, \bibinfo {author} {\bibfnamefont {M.}~\bibnamefont {Kawasaki}}, \
  and\ \bibinfo {author} {\bibfnamefont {M.}~\bibnamefont {Yamazaki}},\ }\href
  {\doibase 10.1103/PhysRevD.98.043531} {\bibfield  {journal} {\bibinfo
  {journal} {Phys. Rev.}\ }\textbf {\bibinfo {volume} {D98}},\ \bibinfo {pages}
  {043531} (\bibinfo {year} {2018})},\ \Eprint
  {http://arxiv.org/abs/1711.10496} {arXiv:1711.10496 [astro-ph.CO]}
  \BibitemShut {NoStop}%
\bibitem [{\citenamefont {Zhou}\ \emph {et~al.}(2013)\citenamefont {Zhou},
  \citenamefont {Copeland}, \citenamefont {Easther}, \citenamefont {Finkel},
  \citenamefont {Mou},\ and\ \citenamefont {Saffin}}]{Zhou:2013tsa}%
  \BibitemOpen
  \bibfield  {author} {\bibinfo {author} {\bibfnamefont {S.-Y.}\ \bibnamefont
  {Zhou}}, \bibinfo {author} {\bibfnamefont {E.~J.}\ \bibnamefont {Copeland}},
  \bibinfo {author} {\bibfnamefont {R.}~\bibnamefont {Easther}}, \bibinfo
  {author} {\bibfnamefont {H.}~\bibnamefont {Finkel}}, \bibinfo {author}
  {\bibfnamefont {Z.-G.}\ \bibnamefont {Mou}}, \ and\ \bibinfo {author}
  {\bibfnamefont {P.~M.}\ \bibnamefont {Saffin}},\ }\href {\doibase
  10.1007/JHEP10(2013)026} {\bibfield  {journal} {\bibinfo  {journal} {JHEP}\
  }\textbf {\bibinfo {volume} {10}},\ \bibinfo {pages} {026} (\bibinfo {year}
  {2013})},\ \Eprint {http://arxiv.org/abs/1304.6094} {arXiv:1304.6094
  [astro-ph.CO]} \BibitemShut {NoStop}%
\bibitem [{\citenamefont {Antusch}\ \emph {et~al.}(2017)\citenamefont
  {Antusch}, \citenamefont {Cefala},\ and\ \citenamefont
  {Orani}}]{Antusch:2016con}%
  \BibitemOpen
  \bibfield  {author} {\bibinfo {author} {\bibfnamefont {S.}~\bibnamefont
  {Antusch}}, \bibinfo {author} {\bibfnamefont {F.}~\bibnamefont {Cefala}}, \
  and\ \bibinfo {author} {\bibfnamefont {S.}~\bibnamefont {Orani}},\ }\href
  {\doibase 10.1103/PhysRevLett.120.219901, 10.1103/PhysRevLett.118.011303}
  {\bibfield  {journal} {\bibinfo  {journal} {Phys. Rev. Lett.}\ }\textbf
  {\bibinfo {volume} {118}},\ \bibinfo {pages} {011303} (\bibinfo {year}
  {2017})},\ \bibinfo {note} {[Erratum: Phys. Rev.
  Lett.120,no.21,219901(2018)]},\ \Eprint {http://arxiv.org/abs/1607.01314}
  {arXiv:1607.01314 [astro-ph.CO]} \BibitemShut {NoStop}%
\bibitem [{\citenamefont {Kolb}\ and\ \citenamefont
  {Tkachev}(1993)}]{Kolb:1993zz}%
  \BibitemOpen
  \bibfield  {author} {\bibinfo {author} {\bibfnamefont {E.~W.}\ \bibnamefont
  {Kolb}}\ and\ \bibinfo {author} {\bibfnamefont {I.~I.}\ \bibnamefont
  {Tkachev}},\ }\href {\doibase 10.1103/PhysRevLett.71.3051} {\bibfield
  {journal} {\bibinfo  {journal} {Phys. Rev. Lett.}\ }\textbf {\bibinfo
  {volume} {71}},\ \bibinfo {pages} {3051} (\bibinfo {year} {1993})},\ \Eprint
  {http://arxiv.org/abs/hep-ph/9303313} {arXiv:hep-ph/9303313 [hep-ph]}
  \BibitemShut {NoStop}%
\bibitem [{\citenamefont {Kolb}\ and\ \citenamefont
  {Tkachev}(1994)}]{Kolb:1993hw}%
  \BibitemOpen
  \bibfield  {author} {\bibinfo {author} {\bibfnamefont {E.~W.}\ \bibnamefont
  {Kolb}}\ and\ \bibinfo {author} {\bibfnamefont {I.~I.}\ \bibnamefont
  {Tkachev}},\ }\href {\doibase 10.1103/PhysRevD.49.5040} {\bibfield  {journal}
  {\bibinfo  {journal} {Phys. Rev.}\ }\textbf {\bibinfo {volume} {D49}},\
  \bibinfo {pages} {5040} (\bibinfo {year} {1994})},\ \Eprint
  {http://arxiv.org/abs/astro-ph/9311037} {arXiv:astro-ph/9311037 [astro-ph]}
  \BibitemShut {NoStop}%
\bibitem [{\citenamefont {Visinelli}\ \emph {et~al.}(2018)\citenamefont
  {Visinelli}, \citenamefont {Baum}, \citenamefont {Redondo}, \citenamefont
  {Freese},\ and\ \citenamefont {Wilczek}}]{Visinelli:2017ooc}%
  \BibitemOpen
  \bibfield  {author} {\bibinfo {author} {\bibfnamefont {L.}~\bibnamefont
  {Visinelli}}, \bibinfo {author} {\bibfnamefont {S.}~\bibnamefont {Baum}},
  \bibinfo {author} {\bibfnamefont {J.}~\bibnamefont {Redondo}}, \bibinfo
  {author} {\bibfnamefont {K.}~\bibnamefont {Freese}}, \ and\ \bibinfo {author}
  {\bibfnamefont {F.}~\bibnamefont {Wilczek}},\ }\href {\doibase
  10.1016/j.physletb.2017.12.010} {\bibfield  {journal} {\bibinfo  {journal}
  {Phys. Lett.}\ }\textbf {\bibinfo {volume} {B777}},\ \bibinfo {pages} {64}
  (\bibinfo {year} {2018})},\ \Eprint {http://arxiv.org/abs/1710.08910}
  {arXiv:1710.08910 [astro-ph.CO]} \BibitemShut {NoStop}%
\bibitem [{\citenamefont {Vaquero}\ \emph {et~al.}(2018)\citenamefont
  {Vaquero}, \citenamefont {Redondo},\ and\ \citenamefont
  {Stadler}}]{Vaquero:2018tib}%
  \BibitemOpen
  \bibfield  {author} {\bibinfo {author} {\bibfnamefont {A.}~\bibnamefont
  {Vaquero}}, \bibinfo {author} {\bibfnamefont {J.}~\bibnamefont {Redondo}}, \
  and\ \bibinfo {author} {\bibfnamefont {J.}~\bibnamefont {Stadler}},\
  }\href@noop {} {\  (\bibinfo {year} {2018})},\ \Eprint
  {http://arxiv.org/abs/1809.09241} {arXiv:1809.09241 [astro-ph.CO]}
  \BibitemShut {NoStop}%
\bibitem [{\citenamefont {Weinberg}(1978)}]{Weinberg:1977ma}%
  \BibitemOpen
  \bibfield  {author} {\bibinfo {author} {\bibfnamefont {S.}~\bibnamefont
  {Weinberg}},\ }\href {\doibase 10.1103/PhysRevLett.40.223} {\bibfield
  {journal} {\bibinfo  {journal} {Phys. Rev. Lett.}\ }\textbf {\bibinfo
  {volume} {40}},\ \bibinfo {pages} {223} (\bibinfo {year} {1978})}\BibitemShut
  {NoStop}%
\bibitem [{\citenamefont {Wilczek}(1978)}]{Wilczek:1977pj}%
  \BibitemOpen
  \bibfield  {author} {\bibinfo {author} {\bibfnamefont {F.}~\bibnamefont
  {Wilczek}},\ }\href {\doibase 10.1103/PhysRevLett.40.279} {\bibfield
  {journal} {\bibinfo  {journal} {Phys. Rev. Lett.}\ }\textbf {\bibinfo
  {volume} {40}},\ \bibinfo {pages} {279} (\bibinfo {year} {1978})}\BibitemShut
  {NoStop}%
\bibitem [{\citenamefont {Kim}(1979)}]{Kim:1979if}%
  \BibitemOpen
  \bibfield  {author} {\bibinfo {author} {\bibfnamefont {J.~E.}\ \bibnamefont
  {Kim}},\ }\href {\doibase 10.1103/PhysRevLett.43.103} {\bibfield  {journal}
  {\bibinfo  {journal} {Phys. Rev. Lett.}\ }\textbf {\bibinfo {volume} {43}},\
  \bibinfo {pages} {103} (\bibinfo {year} {1979})}\BibitemShut {NoStop}%
\bibitem [{\citenamefont {Shifman}\ \emph {et~al.}(1980)\citenamefont
  {Shifman}, \citenamefont {Vainshtein},\ and\ \citenamefont
  {Zakharov}}]{Shifman:1979if}%
  \BibitemOpen
  \bibfield  {author} {\bibinfo {author} {\bibfnamefont {M.~A.}\ \bibnamefont
  {Shifman}}, \bibinfo {author} {\bibfnamefont {A.~I.}\ \bibnamefont
  {Vainshtein}}, \ and\ \bibinfo {author} {\bibfnamefont {V.~I.}\ \bibnamefont
  {Zakharov}},\ }\href {\doibase 10.1016/0550-3213(80)90209-6} {\bibfield
  {journal} {\bibinfo  {journal} {Nucl. Phys.}\ }\textbf {\bibinfo {volume}
  {B166}},\ \bibinfo {pages} {493} (\bibinfo {year} {1980})}\BibitemShut
  {NoStop}%
\bibitem [{\citenamefont {Dine}\ \emph {et~al.}(1981)\citenamefont {Dine},
  \citenamefont {Fischler},\ and\ \citenamefont {Srednicki}}]{Dine:1981rt}%
  \BibitemOpen
  \bibfield  {author} {\bibinfo {author} {\bibfnamefont {M.}~\bibnamefont
  {Dine}}, \bibinfo {author} {\bibfnamefont {W.}~\bibnamefont {Fischler}}, \
  and\ \bibinfo {author} {\bibfnamefont {M.}~\bibnamefont {Srednicki}},\ }\href
  {\doibase 10.1016/0370-2693(81)90590-6} {\bibfield  {journal} {\bibinfo
  {journal} {Phys. Lett.}\ }\textbf {\bibinfo {volume} {104B}},\ \bibinfo
  {pages} {199} (\bibinfo {year} {1981})}\BibitemShut {NoStop}%
\bibitem [{\citenamefont {Peccei}\ and\ \citenamefont
  {Quinn}(1977{\natexlab{a}})}]{Peccei:1977ur}%
  \BibitemOpen
  \bibfield  {author} {\bibinfo {author} {\bibfnamefont {R.~D.}\ \bibnamefont
  {Peccei}}\ and\ \bibinfo {author} {\bibfnamefont {H.~R.}\ \bibnamefont
  {Quinn}},\ }\href {\doibase 10.1103/PhysRevD.16.1791} {\bibfield  {journal}
  {\bibinfo  {journal} {Phys. Rev.}\ }\textbf {\bibinfo {volume} {D16}},\
  \bibinfo {pages} {1791} (\bibinfo {year} {1977}{\natexlab{a}})}\BibitemShut
  {NoStop}%
\bibitem [{\citenamefont {Peccei}\ and\ \citenamefont
  {Quinn}(1977{\natexlab{b}})}]{Peccei:1977hh}%
  \BibitemOpen
  \bibfield  {author} {\bibinfo {author} {\bibfnamefont {R.~D.}\ \bibnamefont
  {Peccei}}\ and\ \bibinfo {author} {\bibfnamefont {H.~R.}\ \bibnamefont
  {Quinn}},\ }\href {\doibase 10.1103/PhysRevLett.38.1440} {\bibfield
  {journal} {\bibinfo  {journal} {Phys. Rev. Lett.}\ }\textbf {\bibinfo
  {volume} {38}},\ \bibinfo {pages} {1440} (\bibinfo {year}
  {1977}{\natexlab{b}})}\BibitemShut {NoStop}%
\bibitem [{\citenamefont {'t~Hooft}(1976)}]{tHooft:1976rip}%
  \BibitemOpen
  \bibfield  {author} {\bibinfo {author} {\bibfnamefont {G.}~\bibnamefont
  {'t~Hooft}},\ }\href {\doibase 10.1103/PhysRevLett.37.8} {\bibfield
  {journal} {\bibinfo  {journal} {Phys. Rev. Lett.}\ }\textbf {\bibinfo
  {volume} {37}},\ \bibinfo {pages} {8} (\bibinfo {year} {1976})}\BibitemShut
  {NoStop}%
\bibitem [{\citenamefont {Fodor}\ \emph {et~al.}(2006)\citenamefont {Fodor},
  \citenamefont {Forgacs}, \citenamefont {Grandclement},\ and\ \citenamefont
  {Racz}}]{Fodor:2006zs}%
  \BibitemOpen
  \bibfield  {author} {\bibinfo {author} {\bibfnamefont {G.}~\bibnamefont
  {Fodor}}, \bibinfo {author} {\bibfnamefont {P.}~\bibnamefont {Forgacs}},
  \bibinfo {author} {\bibfnamefont {P.}~\bibnamefont {Grandclement}}, \ and\
  \bibinfo {author} {\bibfnamefont {I.}~\bibnamefont {Racz}},\ }\href {\doibase
  10.1103/PhysRevD.74.124003} {\bibfield  {journal} {\bibinfo  {journal} {Phys.
  Rev.}\ }\textbf {\bibinfo {volume} {D74}},\ \bibinfo {pages} {124003}
  (\bibinfo {year} {2006})},\ \Eprint {http://arxiv.org/abs/hep-th/0609023}
  {arXiv:hep-th/0609023 [hep-th]} \BibitemShut {NoStop}%
\bibitem [{\citenamefont {Fodor}\ \emph
  {et~al.}(2009{\natexlab{a}})\citenamefont {Fodor}, \citenamefont {Forgacs},
  \citenamefont {Horvath},\ and\ \citenamefont {Mezei}}]{Fodor:2008du}%
  \BibitemOpen
  \bibfield  {author} {\bibinfo {author} {\bibfnamefont {G.}~\bibnamefont
  {Fodor}}, \bibinfo {author} {\bibfnamefont {P.}~\bibnamefont {Forgacs}},
  \bibinfo {author} {\bibfnamefont {Z.}~\bibnamefont {Horvath}}, \ and\
  \bibinfo {author} {\bibfnamefont {M.}~\bibnamefont {Mezei}},\ }\href
  {\doibase 10.1103/PhysRevD.79.065002} {\bibfield  {journal} {\bibinfo
  {journal} {Phys. Rev.}\ }\textbf {\bibinfo {volume} {D79}},\ \bibinfo {pages}
  {065002} (\bibinfo {year} {2009}{\natexlab{a}})},\ \Eprint
  {http://arxiv.org/abs/0812.1919} {arXiv:0812.1919 [hep-th]} \BibitemShut
  {NoStop}%
\bibitem [{\citenamefont {Gleiser}\ and\ \citenamefont
  {Sicilia}(2008)}]{Gleiser:2008ty}%
  \BibitemOpen
  \bibfield  {author} {\bibinfo {author} {\bibfnamefont {M.}~\bibnamefont
  {Gleiser}}\ and\ \bibinfo {author} {\bibfnamefont {D.}~\bibnamefont
  {Sicilia}},\ }\href {\doibase 10.1103/PhysRevLett.101.011602} {\bibfield
  {journal} {\bibinfo  {journal} {Phys. Rev. Lett.}\ }\textbf {\bibinfo
  {volume} {101}},\ \bibinfo {pages} {011602} (\bibinfo {year} {2008})},\
  \Eprint {http://arxiv.org/abs/0804.0791} {arXiv:0804.0791 [hep-th]}
  \BibitemShut {NoStop}%
\bibitem [{\citenamefont {Fodor}\ \emph
  {et~al.}(2009{\natexlab{b}})\citenamefont {Fodor}, \citenamefont {Forgacs},
  \citenamefont {Horvath},\ and\ \citenamefont {Mezei}}]{Fodor:2009kf}%
  \BibitemOpen
  \bibfield  {author} {\bibinfo {author} {\bibfnamefont {G.}~\bibnamefont
  {Fodor}}, \bibinfo {author} {\bibfnamefont {P.}~\bibnamefont {Forgacs}},
  \bibinfo {author} {\bibfnamefont {Z.}~\bibnamefont {Horvath}}, \ and\
  \bibinfo {author} {\bibfnamefont {M.}~\bibnamefont {Mezei}},\ }\href
  {\doibase 10.1016/j.physletb.2009.03.054} {\bibfield  {journal} {\bibinfo
  {journal} {Phys. Lett.}\ }\textbf {\bibinfo {volume} {B674}},\ \bibinfo
  {pages} {319} (\bibinfo {year} {2009}{\natexlab{b}})},\ \Eprint
  {http://arxiv.org/abs/0903.0953} {arXiv:0903.0953 [hep-th]} \BibitemShut
  {NoStop}%
\bibitem [{\citenamefont {Gleiser}\ and\ \citenamefont
  {Sicilia}(2009)}]{Gleiser:2009ys}%
  \BibitemOpen
  \bibfield  {author} {\bibinfo {author} {\bibfnamefont {M.}~\bibnamefont
  {Gleiser}}\ and\ \bibinfo {author} {\bibfnamefont {D.}~\bibnamefont
  {Sicilia}},\ }\href {\doibase 10.1103/PhysRevD.80.125037} {\bibfield
  {journal} {\bibinfo  {journal} {Phys. Rev.}\ }\textbf {\bibinfo {volume}
  {D80}},\ \bibinfo {pages} {125037} (\bibinfo {year} {2009})},\ \Eprint
  {http://arxiv.org/abs/0910.5922} {arXiv:0910.5922 [hep-th]} \BibitemShut
  {NoStop}%
\bibitem [{\citenamefont {Hertzberg}(2010)}]{Hertzberg:2010yz}%
  \BibitemOpen
  \bibfield  {author} {\bibinfo {author} {\bibfnamefont {M.~P.}\ \bibnamefont
  {Hertzberg}},\ }\href {\doibase 10.1103/PhysRevD.82.045022} {\bibfield
  {journal} {\bibinfo  {journal} {Phys. Rev.}\ }\textbf {\bibinfo {volume}
  {D82}},\ \bibinfo {pages} {045022} (\bibinfo {year} {2010})},\ \Eprint
  {http://arxiv.org/abs/1003.3459} {arXiv:1003.3459 [hep-th]} \BibitemShut
  {NoStop}%
\bibitem [{\citenamefont {Saffin}\ \emph {et~al.}(2014)\citenamefont {Saffin},
  \citenamefont {Tognarelli},\ and\ \citenamefont {Tranberg}}]{Saffin:2014yka}%
  \BibitemOpen
  \bibfield  {author} {\bibinfo {author} {\bibfnamefont {P.~M.}\ \bibnamefont
  {Saffin}}, \bibinfo {author} {\bibfnamefont {P.}~\bibnamefont {Tognarelli}},
  \ and\ \bibinfo {author} {\bibfnamefont {A.}~\bibnamefont {Tranberg}},\
  }\href {\doibase 10.1007/JHEP08(2014)125} {\bibfield  {journal} {\bibinfo
  {journal} {JHEP}\ }\textbf {\bibinfo {volume} {08}},\ \bibinfo {pages} {125}
  (\bibinfo {year} {2014})},\ \Eprint {http://arxiv.org/abs/1401.6168}
  {arXiv:1401.6168 [hep-ph]} \BibitemShut {NoStop}%
\bibitem [{\citenamefont {Kawasaki}\ and\ \citenamefont
  {Yamada}(2014)}]{Kawasaki:2013awa}%
  \BibitemOpen
  \bibfield  {author} {\bibinfo {author} {\bibfnamefont {M.}~\bibnamefont
  {Kawasaki}}\ and\ \bibinfo {author} {\bibfnamefont {M.}~\bibnamefont
  {Yamada}},\ }\href {\doibase 10.1088/1475-7516/2014/02/001} {\bibfield
  {journal} {\bibinfo  {journal} {JCAP}\ }\textbf {\bibinfo {volume} {1402}},\
  \bibinfo {pages} {001} (\bibinfo {year} {2014})},\ \Eprint
  {http://arxiv.org/abs/1311.0985} {arXiv:1311.0985 [hep-ph]} \BibitemShut
  {NoStop}%
\bibitem [{\citenamefont {Mukaida}\ \emph {et~al.}(2017)\citenamefont
  {Mukaida}, \citenamefont {Takimoto},\ and\ \citenamefont
  {Yamada}}]{Mukaida:2016hwd}%
  \BibitemOpen
  \bibfield  {author} {\bibinfo {author} {\bibfnamefont {K.}~\bibnamefont
  {Mukaida}}, \bibinfo {author} {\bibfnamefont {M.}~\bibnamefont {Takimoto}}, \
  and\ \bibinfo {author} {\bibfnamefont {M.}~\bibnamefont {Yamada}},\ }\href
  {\doibase 10.1007/JHEP03(2017)122} {\bibfield  {journal} {\bibinfo  {journal}
  {JHEP}\ }\textbf {\bibinfo {volume} {03}},\ \bibinfo {pages} {122} (\bibinfo
  {year} {2017})},\ \Eprint {http://arxiv.org/abs/1612.07750} {arXiv:1612.07750
  [hep-ph]} \BibitemShut {NoStop}%
\bibitem [{\citenamefont {Eby}\ \emph {et~al.}(2018)\citenamefont {Eby},
  \citenamefont {Mukaida}, \citenamefont {Takimoto}, \citenamefont
  {Wijewardhana},\ and\ \citenamefont {Yamada}}]{Eby:2018ufi}%
  \BibitemOpen
  \bibfield  {author} {\bibinfo {author} {\bibfnamefont {J.}~\bibnamefont
  {Eby}}, \bibinfo {author} {\bibfnamefont {K.}~\bibnamefont {Mukaida}},
  \bibinfo {author} {\bibfnamefont {M.}~\bibnamefont {Takimoto}}, \bibinfo
  {author} {\bibfnamefont {L.~C.~R.}\ \bibnamefont {Wijewardhana}}, \ and\
  \bibinfo {author} {\bibfnamefont {M.}~\bibnamefont {Yamada}},\ }\href@noop {}
  {\  (\bibinfo {year} {2018})},\ \Eprint {http://arxiv.org/abs/1807.09795}
  {arXiv:1807.09795 [hep-ph]} \BibitemShut {NoStop}%
\bibitem [{\citenamefont {Mukaida}()}]{Mukaida:PVC}%
  \BibitemOpen
  \bibfield  {author} {\bibinfo {author} {\bibfnamefont {K.}~\bibnamefont
  {Mukaida}},\ }\href@noop {} {}\bibinfo {howpublished} {private
  communication}\BibitemShut {NoStop}%
\bibitem [{\citenamefont {Salmi}\ and\ \citenamefont
  {Hindmarsh}(2012)}]{Salmi:2012ta}%
  \BibitemOpen
  \bibfield  {author} {\bibinfo {author} {\bibfnamefont {P.}~\bibnamefont
  {Salmi}}\ and\ \bibinfo {author} {\bibfnamefont {M.}~\bibnamefont
  {Hindmarsh}},\ }\href {\doibase 10.1103/PhysRevD.85.085033} {\bibfield
  {journal} {\bibinfo  {journal} {Phys. Rev.}\ }\textbf {\bibinfo {volume}
  {D85}},\ \bibinfo {pages} {085033} (\bibinfo {year} {2012})},\ \Eprint
  {http://arxiv.org/abs/1201.1934} {arXiv:1201.1934 [hep-th]} \BibitemShut
  {NoStop}%
\end{thebibliography}%

\end{document}